# Advancements in Programmable Lipid Nanoparticles: Exploring the Four-Domain Model for Targeted Drug Delivery

Zhaoyu Liu[1, 2], Jingxun Chen[1, 2], Mingkun Xu[3], David H. Gracias[4], Ken-Tye Yong[5, 6, *], Yuanyuan Wei[1, 7, *], Ho-Pui Ho[1, *]

[1] Department of Biomedical Engineering, The Chinese University of Hong Kong, Shatin, Hong Kong SAR, 999077, China. E-mail: wei-yy18@link.cuhk.edu.hk; aaron.ho@cuhk.edu.hk.

[2] Department of Biomedical Engineering, Whiting School of Engineering, Johns Hopkins University, Baltimore, Maryland, 21218, USA.

[3] Guangdong Institute of Intelligence Science and Technology, Hengqin, Zhuhai, 519031, China.

[4] Department of Chemical and Biomolecular Engineering, Johns Hopkins University, Baltimore, Maryland, 21218, USA.

[5] School of Biomedical Engineering, The University of Sydney, Sydney, New South Wales 2006, Australia. E-mail: ken.yong@sydney.edu.au.

[6] The Biophotonics and Mechano-Bioengineering Lab, The University of Sydney, Sydney, New South Wales 2006, Australia.

[7] Department of Neurology, David Geffen School of Medicine, University of California, Los Angeles, California, 90095, USA.

[*]Correspondence:

ken.yong@sydney.edu.au; wei-yy18@link.cuhk.edu.hk; aaron.ho@cuhk.edu.hk.

**Abstract**

Programmable lipid nanoparticles, or LNPs, represent a breakthrough in the realm of targeted drug delivery, offering precise spatiotemporal control essential for the treatment of complex diseases such as cancer and genetic disorders. In order to provide a more modular perspective and a more balanced analysis of the mechanism, this review presents a novel Four-Domain Model that consists of Architecture, Interface, Payload, and Dispersal Domain. We explored the dynamical equilibrium between LNPs components and the surroundings throughout their destiny, from formulation to release. On the basis of this, we delve deep into manufacturing challenges, scalability issues, and regulatory hurdles, associated with the clinical translation of LNP technology. Within the framework focusing on the programmability in each domain, we prioritized patient-centric factors like dosing regimens, administration techniques, and potential consequences. Notably, this review expands to innovative anatomical routes, such as intranasal and intraocular administration, offering a thorough examination of the advantages and disadvantages of each route. We also offered a comprehensive comparison between artificial LNPs and natural exosomes in terms of functionality, biocompatibility, and therapeutic potential. Ultimately, this review highlights the potential of programmable LNPs to evolve into more intelligent, naturally integrated systems, achieving optimal biocompatibility and functionality.

# Table of Contents

# 1   Introduction

Complex diseases like cancer and genetic disorders often exhibit heterogeneous and dynamic characteristics within the body, making uniform drug distribution challenging[1,2]. The treatment of these diseases necessitates precise and effective drug-delivery systems with high spatiotemporal control for maximized therapeutic impact[3,4]. Up to date, delivery loaders such as hydrogels[5], polymeric nanoparticles[6], molecular conjugates[7], viral vectors[8], and lipid-based nanoparticles[9] (LBNPs) are getting widely used. Particularly, LBNPs are increasingly recognized for their ability to protect and deliver sensitive agents like mRNA to target sites, locally or systematically, while maintaining biocompatibility[9,10]. LBNPs allow targeted administration to specified diseased cells or tissues, enhancing therapeutic efficacy via controlled release mechanisms and minimizing systemic side effects. Among the diverse LBNP formulations, lipid nanoparticles (LNPs) are distinguished by their inclusion of ionizable lipids. They incorporate helper lipids, cholesterol, PEGylated lipids, and other functional components into their self-assembled nanoscale structure, just like other LBNPs[11,12]. Their impact has been evidenced by numerous clinical trials[13,14] and emergency use authorizations for SARS-CoV2 mRNA vaccines by Pfizer and Moderna[15–17].

Conceptually, LNPs simulate a "thermodynamic system" where the mass is enclosed within confined spaces under equilibrium circumstances and released when the system is non-equilibrium. In this review, we examine the equilibrium and non-equilibrium states of LNPs as semi-closed systems at different stages of their life cycle. While existing reviews comprehensively covered aspects such as development[9], component[18], application[19], and clinical aspects of LNPs[19], our perspective further integrates structural, environmental, and functional aspects within a thermodynamically informed framework. On this basis, we introduce the Four-Domain Model, comprising the Architecture, Interface, Payload, and Dispersal Domain (**Fig. 1**). This modular framework fills in the gaps in the present studies by dissecting the advancement of LNPs with particular attention on each domain's programmability. We apply the term "programmability" to LNPs drug delivery systems here, even though it was originally used to describe microrobots in computer systems or any intelligent system that reacts to external stimuli or acts on its initiative in a predictable and regulated manner. The term "programmability" in this review refers to nanoparticles' capacity to be engineered to respond to certain environmental cues for focused and regulated drug release. The programmability of LNPs, including response to pH[20,21], temperature[11], or enzymatic activity[22,23], has been hugely advanced by developments in materials science and nanotechnology. This advancement has increased therapeutic efficacy and reduced off-target consequences.

In the Four-Domain Model, the Architecture Domain subjects the structural components of LNPs, analogous to the container in a thermodynamic system that separates the payload from the surroundings. This domain's programmability was mostly concentrated on the mesophase structure, which reacts to pH variations and fusion mechanisms in endosome escape. The Interface Domain investigates the interactions between LNPs and their surroundings, focusing on how different functional components respond to environmental factors such as pH, redox potential, and enzymes. The Payload Domain concentrates on the encapsulated mass and its applications, encompassing therapeutic agents and functional payloads with their therapeutic benefits. The Dispersal Domain provides a thorough picture of the lifecycle of LNPs at different stages of delivery by addressing the different administration routes and biodistribution patterns of LNPs. As a result, this architecture provides a modular foundation for the purposeful and clear design of programmable LNPs. "Modularity" refers to the flexibility to add, remove, or modify the components for specific functions or responses, such as targeting ligands or responsive triggers[24–26].

In biomedical contexts, LNPs monitor the environment and maintain equilibrium or close-to-equilibrium before receiving specific cues. The on-demand responsive release, including multi-step delivery[27,28], is conceptualized as a non-equilibrium transition triggered by external stimuli such as pH changes or enzyme presence. By applying these equilibrium descriptions, we gain an intuitive understanding of the dynamic interplay within LNPs. We presented the newest discovery regarding the structure of LNPs by examining manufacturing and regulatory barriers, offering practical recommendations for overcoming these obstacles, and advancing toward clinical adoption. We also suggested potential anatomical delivery routes, such as intracerebral and intraocular methods, that are associated with their challenges and opportunities. Furthermore, we examined And prospected for future research directions to maximize LNPs' potential in precision medicine for the ultimate clinical application. Despite progress on programmable LNPs, natural cells or biological vesicles are non-equilibrium compartments that are far more sophisticated than those in artificial compartments, maintaining intricately interlinked mass and information fluxes. This review further demonstrated a detailed comparison between artificial LNPs and natural exosomes using our Four Domain Model to demonstrate our model's advantage. This comparative analysis provides a roadmap for developing more intelligent, naturally integrated systems. Although LNPs are the review's main focus, we believe these ideas can be expanded to different delivery systems and medical applications to address unmet medical requirements and provide creative answers to challenging health problems.

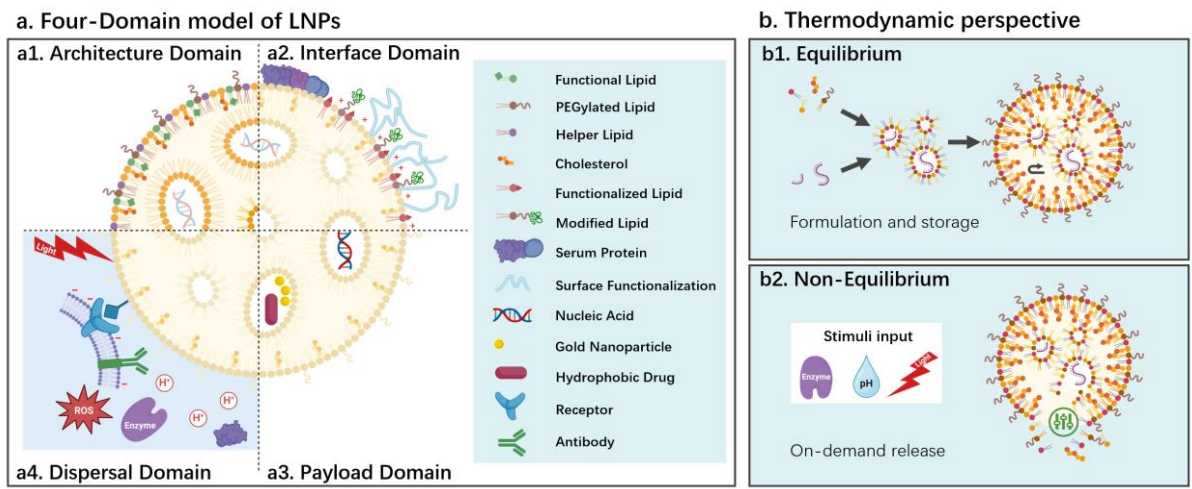

**Fig. 1 Illustration of the Four-Domain Model and thermodynamic perspective of programmable LNPs**. **a.** Schematic overview of the Four-Domain Model of LNPs. **a1.** The Architecture Domain focuses on the structural integrity and composition of LNPs, including ionizable lipids, PEGylated lipids, helper lipids, and cholesterol. **a2.** The Interface Domain addresses interactions with the biological environment, enhancing stability, reducing immunogenicity, and improving targeting efficacy. This domain is the focus of the programmability of LNPs. **a3.** The payload Domain highlights encapsulated therapeutic agents and functional components, focusing on therapeutic effect and programmability. **a4.** The Dispersal Domain explores administration route, biodistribution, and trafficking analysis, which are critical for optimizing delivery strategies and clinical outcomes. **b.** Schematic illustrating LNPs to a semi-closed thermodynamic system in equilibrium (**b1**) and non-equilibrium (**b2**) conditions. **b1.** LNP formulation to achieve equilibrium during storage and delivery. **b2.** LNPs transition to a non-equilibrium state programmed to respond to external stimuli, enabling controlled and on-demand release of payloads.

## 2    Architecture Domain: Structure and Formulation of LNPs

The Architecture Domain is analogous to the container in a thermodynamic system, encapsulating the payload to separate it from the environment. It ensures that the LNPs maintain stability throughout the delivery process, protecting the payload from degradation until it reaches the target site. Architecture Domain comprises the majority of LNPs by molar fraction and primarily consists of cationic ionizable lipids (CILs), PEGylated lipids, helper lipids, and cholesterol, each serving distinct functional roles. These components are self-assembled to form a core-shell architecture that encapsulates and delivers therapeutic payloads while influencing the stability of LNPs and interaction with biological environments[9,19].

### 2.1    Structural Dynamics and Programmability of Architecture Domain: pH-Responsive Behavior and Fusion Mechanisms in the Architecture Domain

The structure of LNPs contains two distinct parts: the shell and the core[29,30]. The shell is a single lipid layer enveloping the LNPs, with its hydrophilic phase facing the exterior to separate the inner and outer environments. The shell component mainly comprises helper lipids, PEGylated lipids, and cholesterol, with only a small portion of CILs[31]. Helper lipids comprise the largest proportion of the LNPs composition and are the primary structural elements. PEGylated lipids are essential for preventing the aggregation of lipid crystalline structures, thereby facilitating the formation of small, discrete nanoparticles. This property is particularly important for maintaining the colloidal stability of LNPs in physiological environments[19]. Cholesterol, another vital component, can modulate the fluidity of lipid membranes, improve their resistance to serum-induced degradation[32], and facilitate endosome escape[29]. The core composition of lipid nanoparticles primarily includes CILs[11], which have largely replaced traditional permanent cationic lipids in liposomal formulations. The inclusion of CIL has significantly reduced biotoxicity and inflammation[19]. Additionally, small amounts of helper lipids and cholesterol are still present in the core[31].

Recent studies suggest that the core is characterized by lipid mesophases that encapsulate aqueous content, forming complex structures such as inverse micellar, inverse hexagonal, cubic, or lamellar phases[33]. The various mesophases that LNPs core can adopt, while not fully understood, are largely governed by the constituent lipids' critical packing parameter (CPP), particularly the CILs and helper lipids lipids[34]. The CPP, defined as the ratio of the amphiphile's hydrophobic molecular volume to the product of its hydrophobic tail length and effective headgroup area (**Fig. 3a**), serves as a predictive tool for lipid self-assembly behavior. A CPP value close to unity typically results in lamellar phase formation. In contrast, values less than one tend to produce inverse hexagonal phase or cubic phase, where the payload is encapsulated within the aqueous center of these inverse structures[35–37] (**Fig. 3b**). This relationship between molecular geometry and supramolecular organization underpins the engineering of programmability into the Architecture Domain of LNPs, which allows us to tune the structure by tuning CPP.

The nature of containing CIL leads to the pH responsiveness of the core. Protonation or deprotonation of these CILs can affect the electrostatic forces between their heads, indirectly affecting the CPP, which results in a change in their structure and mesophase[33]. For instance, at endosomal pH (4), LNPs demonstrate increased volume and core water content compared to physiological pH (7.4), accompanied by disruption of distinct mult-layer shell structure[38]. These pH-induced structural transitions are further exemplified by the mesophase shift from micellar to cubic structures with increasing protonation of CILs in follow-up experiments[39]. Such transitions underpin the inherent programmability without the need to incorporate extra functional components, which are utilized in small molecule release through diffusion and promoting endosomal escape (**Fig. 3c**).

Site-specific delivery of small molecules, such as fluconazole for antifungal therapy[40], can take advantage of this characteristic. Fluconazole can be released by diffusion through water channels in the cubic phase after reaching the acidic environment[40]. Endosomal escape is a limiting step in LNPs delivery. After cellular intake, the LNPs are encapsulated in the endosome and will be degraded when the endosome develops into a lysosome. Escape is essential as therapeutics such as nucleic acids need to reach the cytosol to perform the effect[41]. One primary mechanism of endosome escape is through membrane fusion, where the pH-responsive structural dynamics of LNPs are particularly important[42]. The fusion process is profoundly influenced by the LNP's nanostructure, which can be explained by calculating the energy difference between before and after fusion. The elastic energy can be determined by the Canham-Helfrich formula, $E/A = \frac{\kappa}{2}(J - C_0)^2 + \bar{\kappa}K$, where $E$ is the elastic energy of a membrane and $A$ is the area, $\kappa$ is the bending rigidity, $J$ is the mean curvature and $C_0$ is the intrinsic curvature, $\bar{\kappa}$ is the Gaussian curvature modulus, and $K$ is the Gaussian curvature, which is the product of the two principal curvatures. If we calculate the energy before and after two vesicle fusion, we can see $\Delta E_{fusion} = -4\pi\bar{\kappa}$, which depends on the gaussian modulus. Therefore, a positive Gaussian curvature modulus would promote membrane fusion. While most biology membranes have negative $\bar{\kappa}$, cubic mesophase has a positive $\bar{\kappa}$, which aligns with it being stable with a saddle structure with negative Gaussian curvature $K$. Therefore, compared to laminar or micellar phases, the cubic phase achieves higher fusogenic potential compared to other phases[43,44](**Fig. 3d**). An alternative experiment posits that the insertion of ionizable lipids into the endosome membrane can lead to a hexagonal transformation of the endosome membrane itself[45], facilitating the escape of the payload from the endosome, which can be possibly explained by altering of $\bar{\kappa}$ of the endosome membrane. This intrinsic dynamic of LNPs reveals the principle behind it and can potentially inspire researchers to leverage this characteristic for other uses and invent more novel endosome escape techniques.

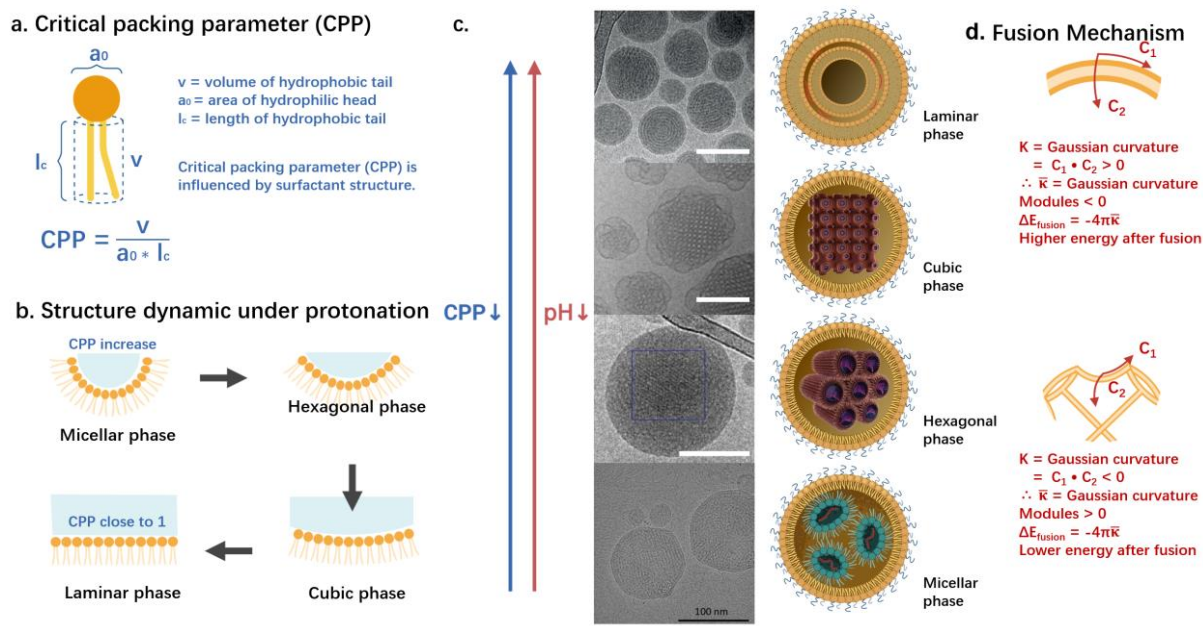

**Fig. 2 Structural of LNPs and mechanisms of endosomal escape. a.** The CPP is a key determinant of the self-assembly structure of surfactants. **b.** A demonstration of how CPP affects the membrane curvature, with blue the hydrophilic phase. **c.** This panel depicts four distinct internal phases of LNPs with Cryo-TEM image: cubic, hexagonal, micellar, and lamellar structures, which CPP and environmental pH levels influence. Changes in CPP and pH drive transitions between these phases, as indicated by the arrows. All scale bars are 100 nm. Cubic and hexagonal phases adapted from ref. [29] with permission. © 2023 The Authors. Published by Elsevier Ltd. **c.** This panel explains how different phases affect the energy during membrane fusion.

## 2.2 Formulation, Scalability, Stability,and Characterization

The core-shell architecture of LNPs is formulated through advanced formulation techniques that have significantly evolved in recent years [46,47]. Traditional methods[48], like thin-film hydration, microemulsion, and solvent injection, have been largely supplanted by more simple and efficient approaches (**Table 1**). Notably, microfluidic technologies have revolutionized the field, offering improved control over particle size, polydispersity, and encapsulation efficiency[49–51] by leveraging laminar flow for precise mixing at the microscale, which is fundamentally governed by thermodynamic and fluid dynamic principles[46,52].

Moreover, the rapid mixing kinetics in microfluidic systems, occurring on millisecond timescales, enable the encapsulation of labile molecules under mild conditions, preserving their integrity. This process involves the rapid mixing of lipids dissolved in an organic phase with an aqueous phase containing nucleic acids, inducing supersaturation and leading to spontaneous nucleation and growth of LNPs encapsulating the payload.

The scalability of LNP production remains a critical concern as these nanoparticles transition from laboratory-scale synthesis to industrial-scale manufacturing for clinical applications. Microfluidic technologies have progressed in addressing this challenge through the principle of parallelization, wherein multiple microfluidic units operate simultaneously to increase production volume while maintaining precise control over particle characteristics achieved at smaller scales[53].

During storage, LNPs face potential degradation pathways, including aggregation[54,55], lipid oxidation[56], payload leakage[57], and payload degradation[58]. The composition of the core is carefully engineered to mitigate these issues[18]. PEG chains, typically incorporated as PEG-lipid conjugates, serve as primary stabilizing agents by creating a steric barrier around the LNP[59]. This PEG-mediated stabilization involves both enthalpic and entropic contributions, with hydrated PEG chains forming a localized aqueous phase and configurational entropy resisting compression when particles approach each other. Other components of the core also contribute to storage stability. Cholesterol enhances structural stability, which is particularly important for maintaining LNPs effectiveness[32]. Lyophilization (freeze-drying) is often employed to enhance the long-term stability of LNPs during storage. However, this process can induce stress on the nanoparticles, potentially leading to aggregation or structural changes. To mitigate the effects of lyophilization, cryoprotectants such as trehalose or sucrose are often added to the formulation to prevent fusion and maintain the structural integrity of LNPs during the freeze-drying process[60].

The structure of LNPs can be characterized with microscopic and scattering techniques. Cryogenic transmission electron microscopy (cryo-TEM) provides a direct way to visualize, offering high-resolution imaging of LNPs, revealing intricate details of their internal mesophase, organization, and morphology[61]. Complementing direct visualization, small-angle X-ray scattering (SAXS) [61] and small-angle neutron scattering (SANS)[62] provide periodic arrangements of lipid molecules in various mesophases, helping us characterize the LNPs' mesophase. Synchrotron-based SAXS, in particular, offers superior resolution and the capability for time-resolved studies, capturing dynamic structural changes during LNP formation and payload release[62]. Complementing experimental approaches, computational simulation techniques, as summarized in **Table 2**, offer advantages by passing wet lab, enabling atom-level kinetic studying, and potential to be scaled with more computation units (CPUs and GPUs). Generally, developing computational simulation techniques for LNPs aims to optimize their design to predict behavior in physiological conditions while reducing the time and cost of drug development. However, computation power can also be a limiting factor, as complex computation models for simulation, especially when ultizing neural networks[63], can require high-end computational units while only simulating hundreds of atoms, which is not directly applicable to LNPs with many more molecules.

.

## 3 Interface Domain: Key to Programmability

The Interface Domain represents the boundary between the LNPs and their surrounding environment, which is analogous to the interface in a thermodynamic system. It encompasses the surface properties, such as charge, affinity to plasma proteins, and ligand presentation, which govern the interaction of the LNPs with biological membranes and other cellular components. Comprising the outermost layer of the LNP, this domain includes polyethylene glycol (PEG) coatings, surface functionalization, and various modifications that collectively determine the particle's *in vivo* fate[19].

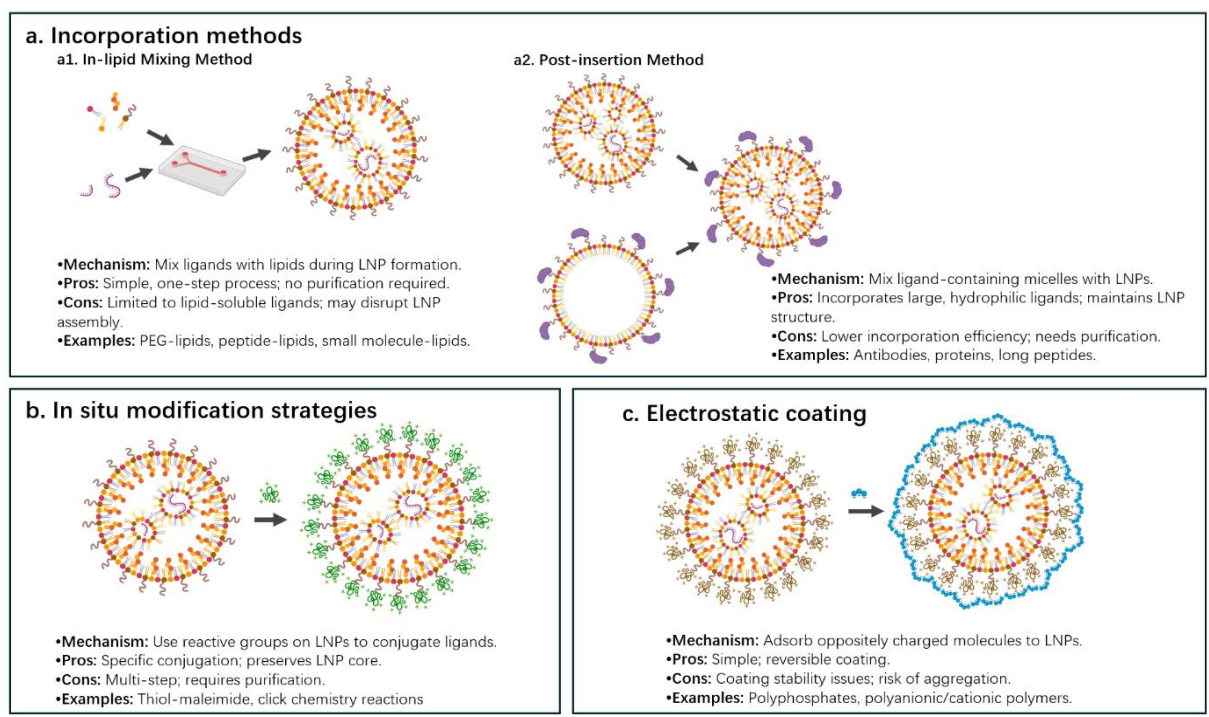

**Fig. 3 Surface engineering of the interface domain. a.** Incorporation methods. Techniques for integrating functional molecules directly into nanoparticles during lipid molecule synthesis. This approach involves pre-synthesizing the nanoparticles with desired molecules to achieve surface engineering. **b.** *in situ* modification strategies. Approaches for modifying nanoparticle surfaces post-synthesis include conjugating targeting ligands or other functional groups via covalent bonds. **c.** Electrostatic coating. Utilizes electrostatic interactions to coat nanoparticles with charged molecules, enhancing stability and providing additional functionalization options. Adapted from ref.[64] with permission. © 2023 The Author(s). Published by Elsevier B.V.

### 3.1 Component of Interface Domain and Surface Engineering Technique

The Interface Domain of LNPs can achieve programmability by integrating various kinds of function components on the surface, including small molecules, targeting proteins, nucleic acid aptamers, functional lipids, and other stimuli-responsive elements. This requires sophisticated engineering of the LNPs interface, necessitated by the complexity of biological systems and the need to control particle-environment interactions precisely. Researchers have developed three primary techniques for integrating functional elements into the Interface Domain to achieve surface engineering. Incorporation of pre-synthesized ligands[65], in which lipid molecules with modification are synthesized before incorporating on LNPs. The modified lipid molecules can be integrated into LNPs by introducing them during formulation (**Fig. 3a1**) or by mixing micellar with functional lipids with LNPs to incorporate these lipids (**Fig. 3a2**). *in situ* modification strategies[65] involve modification of the surface of LNPs by forming a covalent bond of the modifier and the surface of LNPs (**Fig. 3b**). Electrostatic coating[64] utilizes electrostatic interactions to coat nanoparticles with charged molecules, enhancing stability and providing additional functionalization options (**Fig. 3c**). Different approaches can be applied for incorporation and in situ methods, which has been extensively covered by previous review[65].

### 3.2 Programmability of Interface Domain

Programmability in the Interface Domain of LNPs is achieved through three primary strategies: active targeting, the incorporation of environment-responsive elements, and endogenous targeting (**Fig. 4**).

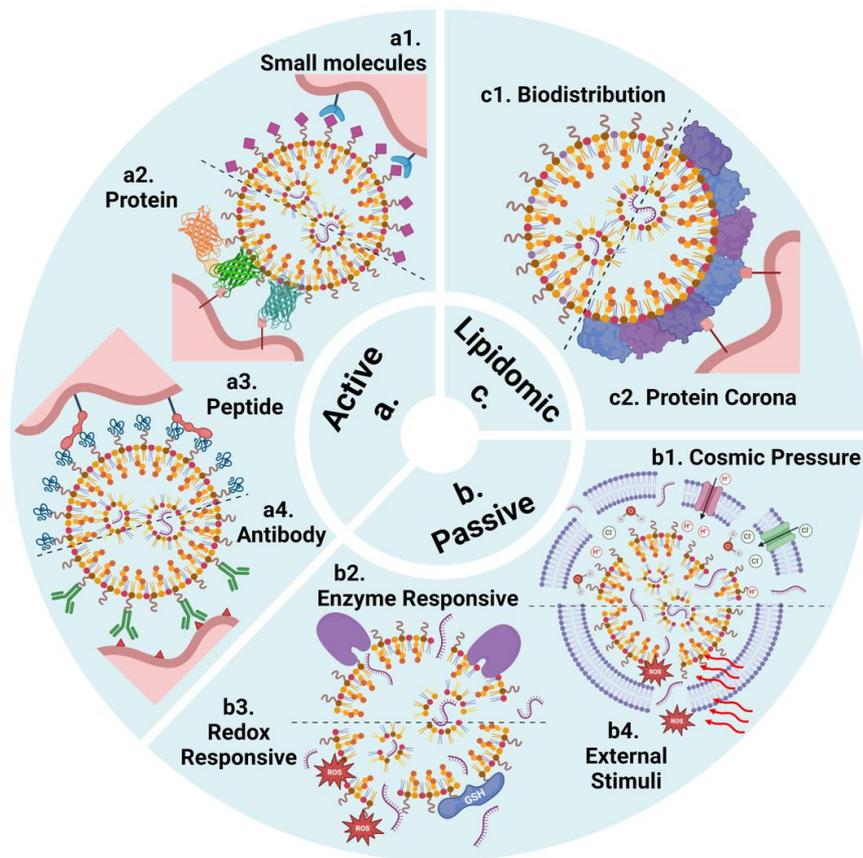

**Fig. 4 Strategies for enhancing programmability in the Interface Domain of LNPs.** Programmability in the Interface Domain of LNPs is achieved through three primary strategies:

a. Active targeting, which involves integrating targeting molecules such as small molecules, peptides, proteins, and nucleic acid aptamers. b. Passible targeting by incorporating environment-responsive elements to respond to various stimuli such as pH changes, redox potential, and enzymatic activity. c. Lipidomic engineering, which can affect the properties of LNPs and the protein corona formation, can affect the targeting ability.

### 3.2.1 Active Targeting: Small Molecules, Peptides, Proteins, and Nucleic Acid Aptamers

As the most popular and researched area, the integration of targeting molecules on the LNPs surface capitalizes on the selectivity of biological interactions to guide nanoparticles to specific cell types or tissues. This strategy employs three main categories of targeting ligands: small molecules, peptides and proteins, and nucleic acid aptamers.

Small molecules can be sugar monomers, nucleoside bases, or even vitamins. They target cancer cells or other specific cells with corresponding receptors on the surface. Folic acid, which is a B vitamin, has been integrated into LNPs as many cancer cells contain folate receptors[66]. Similarly, mannose[67], a sugar monomer, is used to target liver sinusoidal cells with mannose receptors, while adenosine-functionalized LNPs are employed to target astrocyte cells[68].

Peptide ligands can selectively guide LNPs to cells with higher affinity to the peptide. These cells may naturally interact with certain peptides, enabling us to target healthy cells, such as neurons in the neural retina with MH42 peptides-modified LNPs[69]. Sometimes, the cell is only accessible by the peptide under pathological conditions, therefore allowing targeting malfunctioning cells, such as targeting PD-L1-expressing PTEN-deficient triple-negative breast cancer cells with PD-L1 binding peptides decorated LNPs[70]. Antibody modification provides another approach to directly targeting immune cells, with notable progress of Anti-Ly6c antibody modification for targeted delivery to inflammatory leukocytes[71] and Anti-CD5 antibody modification to T cells[72].

Nucleic acid aptamers represent a versatile option, combining the specificity of antibodies with easier production. The effectiveness of aptamers was demonstrated by targeting osteoblasts to achieve gene silencing[73], although the LBNP involved does not perfectly align with our definition. While limited work utilizes this, recent FDA approval of the second aptamers demonstrates an emerging interest and opportunity in this field[74]. In conclusion, these three approaches highlight the versatility and potential of targeted LNPs systems in advancing precision medicine and targeted therapies.

### 3.2.2 Functional Components for Passive Targeting: pH, Redox, Enzymatic, and Light-Triggered Components

Environment-responsive elements incorporated into the Interface Domain enable LNPs to respond dynamically to their surroundings, facilitating controlled payload release. Various stimuli, including pH changes, redox potential, enzymatic activity, and light irradiation, can trigger the responsiveness of these elements.

pH-sensitive element functionalized LNPs focusing on resolving problems related to endosome escape, on top of the programmability in the Architectural Domain introduced earlier. Three primary mechanisms exist: enhancing interaction with the negatively charged membrane after gaining a positive charge in acidic pH, enhancing endosomal escape and cytosolic delivery via the proton sponge effect, and, in a recent study, applying mechanical force to the endosome membrane.

The first approach involves utilizing the pH-activated lipid-like material ssPalm, incorporating vitamin E scaffolds, which destabilizes membranes at low pH through the protonation of tertiary amines[20]. PGlu(DET-Car)30 LNPs, having similar principles, present an active charged surface that interacts with negatively charged membranes in response to pH changes[75]. The proton sponge effect can be achieved by integrating pH buffers, such as using imidazole-based lipids such as O12-D3-I3, which exhibit a buffering effect that induces swelling and enhances endosomal escape[76]. It is also possible to leverage both mechanisms to promote endosome escape, demonstrated by a recently developed acid-degradable anionic lipid LNPs that will undergo hydrolysis of azido-acetal linker under acidic conditions in the endosome and increase the osmolarity while shifting to a positive charge to promote fusion with endosome membrane[77]. The third approach to enhancing endosome escape involves incorporating "nanomachine" lipid components capable of rotation and mechanical transition under a low pH environment, which disrupts the endosome membrane[78].

Redox-responsive elements, including disulfide bonds and thioketal linkages, usually target tumor tissues by exploiting the reducing microenvironment to initiate payload release. Disulfide bonds, commonly used in biologics and subject to specific use case approvals, are cleaved by glutathione (GSH) within the hydrophobic tails of the lipids, leading to truncation and disintegration of the LNPs structure[79]. More advanced designs, such as the BAmP-TK-12 ionizable lipid with a thioketal linkage, respond to elevated reactive oxygen species (ROS) levels in tumor cells[80], where the oxidation and cleavage of the thioketal linkage destabilize the LNPs, triggering payload release. DDA-SS-DMA ionizable lipid contains disulfide bonds that are cleaved in the high GSH environment of tumors, causing nanoparticle disassembly and controlled release of the therapeutic cargo[81].

Enzymatically cleavable linkers are incorporated to treat specific enzymes present locally as input to control the cellular intake or payload release. Existing systems generally go through alterations in the surface charge as enzyme cleavage will remove old groups and induce new groups with different charges. For instance, decationization can be achieved by the corporation of esterase-labile quaternium lipidoid (AMP-POC18)[82], which remain permanently charged until exposed to esterase. Polyphosphate (TPP) incorporated LNPs, which undergo charge reversal triggered by alkaline phosphatase (ALP), can enhance cellular uptake and transfection efficiency at the target site[64]. This change of ionization can also lead to mesophase changes. Phospholipid D (PLD) LNPs with initially cubic structure changed to hexagonal phase after being cleaved by PLD enzymes[23].

Light-responsive elements, such as second near-infrared (NIR-II) dye-conjugated lipids and porphyrin-lipids, offer externally controlled, spatiotemporally precise activation of therapeutic functions through photothermal and ROS-generating mechanisms. NIR-II dye conjugated lipids (Cy-lipids) exhibit increased NIR-II absorption upon protonation in acidic conditions, facilitating photothermal endosomal escape[83]. Similarly, porphyrin-lipids generate ROS upon near-infrared (NIR) light irradiation, disrupting endosomal membranes and promoting payload release[84]. With all these different modalities that take different input stimuli and actuate precise release, we can utilize external stimuli to expand the scope of programmability and potentially achieve intelligence beyond the natural-system level.

### 3.2.3 Lipidomic Engineering of LNPs: Tailoring Physicochemical Properties and Biodistribution for Enhanced Precision in mRNA Delivery.

By adjusting the ratios and types of lipids in the formulation, researchers can modulate the LNP's surface solubility, surface charge, and surface affinity, allowing for fine-tuning the LNP's pharmacokinetics and therapeutic efficacy[18]. The proportion of saturated, unsaturated, and cholesterol lipids can contribute to surface fluidity, affecting the stability of LNPs, and affecting processes such as endosomal escape. Recent insights into lipid saturation reveal that unsaturated lipids enhance mRNA transfection efficiency due to increased fluidity, decreased phase transition temperature, and enhanced fusogenic properties[85]. The surface charge at the physiological condition is determined by the proportion of permanent cationic lipids and ionizable lipids and their pK value. Tuning of surface charge can lead to organ-targeted delivery by affecting their interaction with the immune system in different organs. For instance, negative-charged LNPs are usually used to target the spleen, while positive-charged LNPs can target the lung.

Notable advancements in the tuning of surface affinity include the development of Selective Organ Targeting (SORT) technology, which leverages specific SORT molecules to dramatically alter biodistribution for precise organ targeting[86]. These SORT molecules, which can be permanently cationic (e.g., DOTAP), anionic (e.g., 18PA), or ionizable cationic lipids (e.g., DODAP), dramatically alter the biodistribution of LNPs, enabling precise targeting to organs such as the lungs, spleen, or liver. The type and percentage of SORT molecule used can be finely tuned to achieve the desired tissue tropism, representing a powerful new tool in the lipid-based nanoparticle arsenal. While efficient incorporation of SORT molecules has been a limiting factor, a novel synthesis method of amidine-incorporated degradable (AID) lipids via a streamlined one-pot tandem multi-component reaction (T-MCR) addresses synthesis challenges[87].

Moreover, reassessing traditional LNPs components has shown that helper lipids and cholesterol, once deemed essential, may not be mandatory for LNPs functionality. This paradigm shift is exemplified by the finding that cholesterol removal can address the persistent challenge of preventing nanoparticle accumulation in hepatic tissues, which can be explained by tuning the affinity to reduced lipoprotein adsorption[88]. By modulating and simplifying intrinsic LNPs components, researchers have achieved concurrent mRNA accumulation and translation in targeted organs like the lung and liver. This innovative targeting strategy shows promise for enhancing the precision of mRNA therapies across diverse diseases and may apply to existing LBNPs systems. However, it is important to note that the optimal LNPs formulation likely depends on the specific therapeutic target and delivery requirements, and traditional components may still play crucial roles in various applications. These advancements collectively represent a significant step forward in the development of more precisely targeted and efficient mRNA delivery systems, potentially expanding the therapeutic potential of mRNA-based treatments.

### 3.2.4 Multiple Input Programmability: Leveraging Modularity for LNPs Design

The forthcoming discussion illuminates the transformative potential of LNPs that incorporate multiple responsive elements. LNPs can process complex environmental cues and execute sophisticated, preprogrammed behaviors by incorporating pH-sensitive, redox-responsive, enzymatically cleavable, and light-triggered elements. Works already exist that process more than one input, such as incorporating pH-sensitive NIR-II dye, whose absorption at 1055 nm increased 7.8-fold in acidic conditions (pH 5.0) compared to pH 7.4[83], to achieve photo-thermal effect only at endosome, where low pH exists. It is also to integrate redox-responsive disulfide bonds with pH-activate ssPalm that enhance endosome escape, maximizing therapeutic effects only at tumor sites[20].

In conclusion, the Interface Domain of LNPs represents a dynamic and highly tunable platform for achieving programmable drug delivery. By leveraging advanced engineering techniques and materials, researchers are creating increasingly sophisticated nanocarriers capable of navigating complex biological barriers and responding intelligently to their environment. As our understanding of the interplay between LNPs interfaces and biological systems deepens, we can anticipate developing next-generation programmable LNPs with higher intelligence and close to natural biology systems.

## 4   Payload Domain: Extending Beyond Therapeutic Effects

The Payload Domain corresponds to the therapeutic agent encapsulated within the core of Architecture Domain, akin to the mass in a thermodynamic system. This domain focuses on encapsulated payloads, their clinical application, and their programmable aspects. The encapsulated payloads accommodate diverse therapeutic agents, encompassing both hydrophilic and hydrophobic molecules. Hydrophilic components, typically encapsulated within the aqueous phase, interact electrostatically with the cationic headgroups of the lipids[29]. In contrast, hydrophobic payloads are incorporated within the lipid structure itself, partitioning into the hydrophobic regions of the LNP[89]. This dual capacity for payload incorporation allows LNPs to serve as versatile carriers for various therapeutic agents, from small-molecule drugs to macromolecules such as nucleic acids and proteins (**Table 3**).

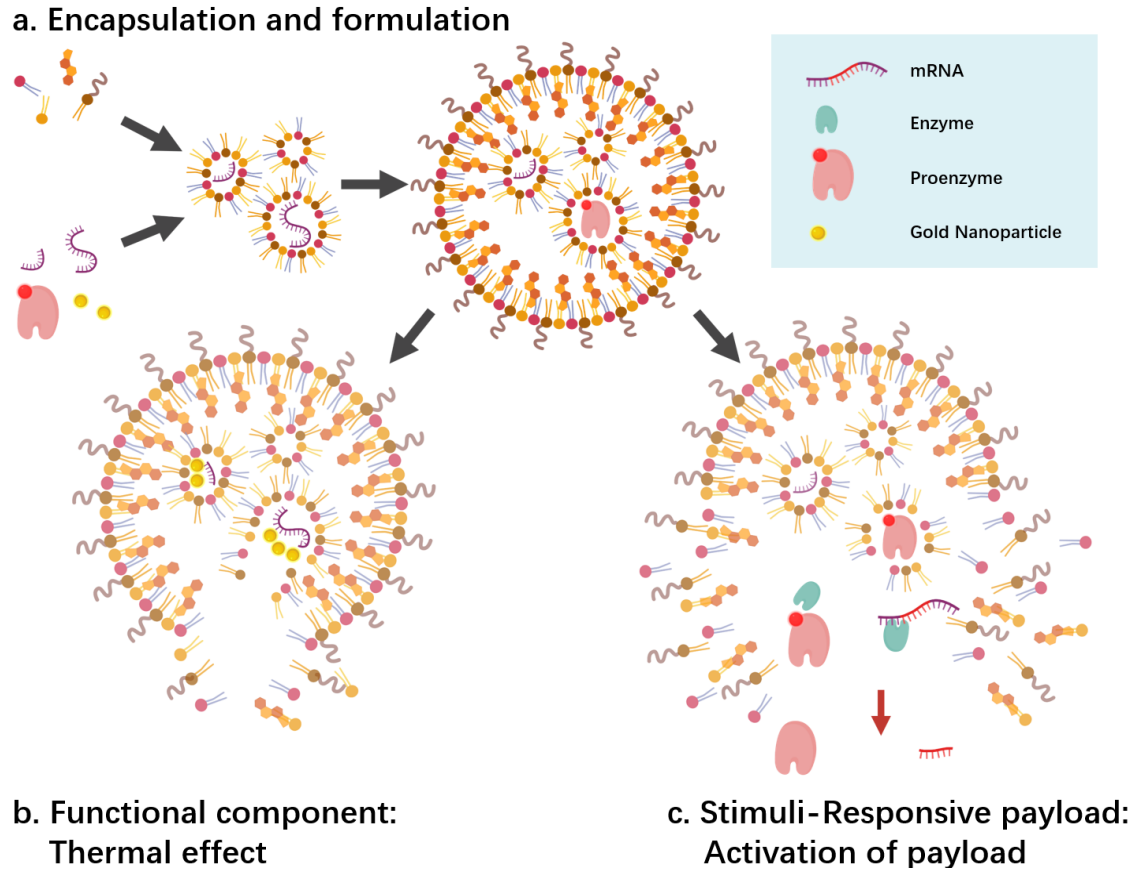

**Fig. 5 Programmability of LNPs utilizing the Payload Domain. a.** The programmability of LNPs can be achieved by encapsulating responsive payloads that react to specific cellular or environmental stimuli. Two primary approaches are highlighted: **b.** Incorporating stimuli-responsive components like gold nanoparticles or photosensitive molecules that trigger payload release upon external stimuli. **c.** Encapsulating enzyme-responsive therapeutic agents that activate selectively in specific intracellular environments. These strategies enhance the precision and efficacy of LNP-based drug delivery systems.

## 4.1 Therapeutic Agent

Contemporary LNP payloads encompass a range of therapeutic agents, each with specific functions and applications. mRNA[13,90] directly delivers genetic instructions for antigen or protein production to cells, finding applications in vaccines, protein replacement therapy, and cancer immunotherapy. saRNA[91,92] allows for intracellular replication and prolonged antigen expression, potentially reducing vaccine doses while enhancing immune response duration. DNA enables sustained antigen production through nuclear transcription and cytoplasmic translation, improving the immunogenicity of DNA vaccines. siRNA[93–95] facilitates gene silencing by downregulating specific genes, with applications in cancer therapy and treating genetic disorders such as hereditary transthyretin-mediated amyloidosis. miRNA[96–98] modulates gene expression by simultaneously targeting multiple genes, offering potential in cancer therapy and treating complex genetic disorders. CRISPR-Cas9[99,100] components enable direct correction of genetic mutations, opening avenues for gene therapy across various genetic conditions. Beyond nucleic acids, LNPs can encapsulate traditional small-molecule drugs with enhanced delivery properties. Chemotherapeutic agents like doxorubicin and paclitaxel benefit from improved solubility, circulation time, and tumor accumulation when delivered via LNPs[101,102]. Antifungal drugs like fluconazole exhibit enhanced delivery and efficacy when formulated in LNPs, improving the treatment of systemic fungal infections[40].

### 4.2 "Debugging" of Programmable LNPs

In programming, achieving accuracy on the first attempt is often challenging. In computer programming, we can always use the 'print()' function to track the change of a particular variable in the terminal. In the context of programmable LNPs, the payload is the 'print()' function, and various imaging techniques will be the terminal. The payload can be engineered to help access the pharmacokinetics (PK) and the pharmacodynamics (PD) of the LNPs, particularly in the context of absorption and distribution.

PK is the study of how the body affects LNPs after administration, and the Payload Domain can be used to access the absorption and distribution of LNPs. Absorption is how the LNPs reach from the administration site into the bloodstream, which can be accessed by analyzing the content in the blood. This can be achieved by detecting the payload mRNA concentration. This can also performed by introducing radioactive isotopes such as $^3$H to track the presence in blood[103]. Distribution is how LNPs spread throughout the body and how they behave during cellular intake, and this intricate process can be tracked by integrating fluorophores. Imaging techniques such as fluorescence imaging and bioluminescence imaging can be used to estimate this process, even in a real-time manner[104]. Formulation of protein corona is also an important process during distribution. However, accessing protein corona is challenging due to the difficulty of separating LNPs from unbounded plasma proteins. The payload can be tailored to solve this problem by integrating superparamagnetic iron oxide nanoparticles (IONPs) in their core and separating them with the magnetic separation method[105].

PD is the study of the effects of the payload of LNPs after delivery, including the subsequent protein expression and biological effects. This can be studied by using a payload that leads to the expression of bioluminescent proteins, such as luciferase, to measure the expression level of proteins in different conditions to model the link between LNPs design and their PD[103].

### 4.3 Programmability of the Payload Domain: Tailoring Therapeutic Activation through External and Intracellular Stimuli

The versatility of payload encapsulation in LNPs has unveiled new horizons for programmability by integrating inherently responsive payloads designed to react to specific cellular or environmental stimuli (**Fig. 5a**). Programmability in the Payload Domain of LNPs can be achieved through two primary approaches. The first approach involves encapsulating stimuli-responsive functional components, such as gold nanoparticles or photosensitive molecules, within the LNPs structure (**Fig. 5b**). These components act as nanoscale transducers, converting external stimuli like light or heat into localized physical or chemical changes, triggering payload release or activation through energy transduction mechanisms[106]. However, the encapsulated interactions between functional components and therapeutic agents should be carefully examined. The second approach focuses on encapsulating enzyme-responsive therapeutic agents, which are activated selectively in specific intracellular environments (**Fig. 5c**). For instance, NQO1-responsive proenzymes represent an innovative strategy wherein the therapeutic payload is activated exclusively by enzymes present in the tumor cell cytosol, thereby achieving a high degree of specificity in anti-cancer treatments[22]. Similarly, another work designed IL-2F, a construct comprising interleukin-2 (IL-2), a cleavable linker, and CD25, as the payload[107]. This payload is engineered to be cleaved by matrix metalloproteinases (MMPs) prevalent in the tumor microenvironment, resulting in the selective release of active IL-2 within cancer tissues. These approaches achieve programming of LNPs with certain intracellular environments as input stimuli while being completed solely by incorporating stimuli responsive payloads. This marks the potential for complex programmable LNPs.

## 5 Dispersal Domain: Administration, Extracellular, and Intracellular Trafficking

The Dispersal Domain is the external milieu that interacts with the system and influences its behavior through cross-border exchanges of matter and energy, similar to the surroundings of a thermodynamic system. Similar to the dispersal of mass in a thermodynamic system, the Dispersal Domain is concerned with the lifetime of LNPs *in vivo*. This domain involves the administration route of LNPs as well as the post-injection dynamics of LNPs.

### 5.1 Administration Route and Extracellular Trafficking of LNPs

The administration route of LNPs determines their site of action, biodistribution, and subsequent biological response. This route influences not only the physiological milieu in which the LNPs are introduced but also the therapeutic efficacy and safety profile of the delivered payload. By selecting and understanding the administration routes, researchers can precisely identify the requisite stimuli for programmable LNPs, thereby optimizing their function to achieve targeted therapeutic delivery. This strategic approach ensures that the LNPs interact effectively with the intended biological targets, minimizing off-target effects and enhancing overall therapeutic outcomes (**Fig. 6**).

#### 5.1.1 Parenteral Route

The parenteral route involves the direct injection of LNPs into the body, bypassing the gastrointestinal tract. This approach, the most widely used method, can be classified into several subcategories based on the site of administration, each with distinct characteristics and implications for LNPs' biodistribution, therapeutic efficacy, and patient experience. The intramuscular (IM) route is particularly favored within this category, especially for SARS-CoV-2 mRNA vaccines. This method introduces local immune responses by delivering the vaccine directly into muscle tissue. In this case, prolonged retention at the injection site will exist[108], and the immune response will be triggered at the lymphatic nodes, as LNPs travel through the lymphatic system without being affected by anti-PEG Antibodies[109]. A recent study with a mouse model investigating the dynamic process between the IM route and the intravenous (IV) route observed that after the IM administration, partial LNP was taken up by muscle cells right after administration and movement of LNPs into the lymphatic system, and bloodstream afterward. The LNPs are transported to the liver within 3 hours of injection, and the protein expression was observed within 1 hour and 3 hours at the injection site and the liver, respectively[110].

The IV route is achieved by directly injecting LNPs into the circulation system, achieving the highest bioactivity[111]. It is widely studied and used by Doxil and Onpattro, with different pharmacokinetics compared to the IM route. After injection, LNPs rapidly accumulate in the liver, as found in the same comparison study[110]. The protein expression was detectable after 1 hour from the liver. It peaked at 3 hours while substantially reducing after 3 hours, indicating a more rapid therapeutic effect, which might result from the direct administration without retention as in the IM route. This characteristic makes the IV route suitable for applications that need rapid and short response, with a higher need for drug concentration.

Subcutaneous (SC) injections are injecting drugs between the skin and muscle, which do not require special personnel and can be completed by patients. Furthermore, SC injections are less painful, have a lower infection rate, and, in rare cases, lead to systemic infection[112]. Similar to IM injection, SC also requires an absorption process after administration. A comparison between SC injection and IV injection demonstrates that while 97.6% of the protein was expressed in the liver hepatocytes in IV injection, around 99% of expression was in the local injection site in the case of SC injection, with predominate expression in adipocytes, supplemented by fibroblasts and macrophages. The plasma concentration for the express protein is 20-fold lower for SC injection than IV injection, while it has a more prolonged effect when applied at the same dose[113]. Therefore, it is critical to consider the pharmacokinetic profile when designing LNPs that potentially require self-administration.

Intradermal (ID) injections are defined as the administration of a drug into the dermis, which is the layer just below the outmost layer, with a slight needle angle. While not commonly used in LNPs administration, it is getting attention as the skin is easily accessible while immune competent[114]. After ID injection, the expression can be observed at the injection site and near the lymph node in a mouse experiment, which can stem from the traveling of either transfected antigen-presenting cells (APCs) or LNPs[115].

Specialized forms of parenteral administration have also been investigated in LNPs delivery. The intracerebral injection has been developed for direct delivery to the brain[116], achieving local transfection and bypassing the blood-brain barrier (BBB) while requiring a highly invasive procedure with associated risks. Local transfection in the eye has also been demonstrated via subretinal injection in mice[117]. The intracerebroventricular injection is another type of parenteral administration where LNPs are delivered into the cerebrospinal fluid (CSF) within the brain's ventricles. After injection into the lateral ventricles of a mouse model, LNPs or injected substances can spread out with CSF flow, which will then later penetrate into brain tissue. However, LNPs are blocked away from the brain tissue. They are primarily expressed in lining cells of the ventricles and blood vessels, resulting in a lower in vivo transfection rate compared to free oligonucleotides[118]. While this is a promising route, it would require future optimization on the LNPs. A novel approach, the needle-free (NF) injection, is also explored. This method has several advantages, including addressing needle phobia, reducing reliance on trained professionals, improved LNPs distribution, and the versatility to select between IM or SC injection[119]. The injected LNPs remain integrated after NF injection while covering a larger area with a higher antibody response than needle-injected IM injection in a rabbit model[119]. However, a real-time trafficking analysis is necessary to further understand the kinetic after NF injection.

### 5.1.2 Topical Route

The topical Route refers to applying substances directly onto the skin or mucous membranes to achieve a therapeutic effect, which can include eye drops, ear drops, and lotions. This method is preferred if we aim to deliver to a target site non-invasively, especially at the open wound where LNPs can pass through the skin barrier. For instance, applying LNPs encapsulating locked nucleic acids within hydrogels on the burnt wound enhances the expression of proteins essential for skin barrier function. It accelerates wound healing, with repeated administration possible[120]. Eye drops are also a promising administration route due to their non-invasive nature and ease of use. However, the limited efficacy of the tear fluid and ocular barriers prevents this approach for LNPs application[121]. Previous investigations of eye drop with liposomes show that smaller liposomes are easier to travel to the retinal pigment epithelium 5 minutes after eye drop administration, potentially through choroidal vessels[122]. Despite the process, direct trafficking of liposomes may be needed to understand the pharmacokinetics of LBNP administration with eyedrops.

Pulmonary delivery can also be seen as a form of the topical route, which involves delivering LNPs to the surface of pulmonary mucosa via inhalation. This approach achieves non-invasive targeted delivery to the lung and reduces off-target effects in other organs. While challenges such as stabilization of LNPs during nebulization and penetration exist, recent advances have explored the role of PEGylated lipids in addressing these issues[123,124]. Increased stability, enhanced mucus penetration, and elevated alveoli and respiratory bronchioles expression levels have been observed. Moreover, sustained protein production is achieved through repeated administration[123]. Similar to the pulmonary route, the intranasal route involves administration into the nasal cavity and aims to target the nasal mucosa and the upper respiratory track. Similar to the pulmonary route, LNPs need to travel through nasal mucosal for effective delivery, and this process is demonstrated to be facilitated by the incorporation of extra cationic lipids in LNPs[125]. After intranasal administration, LNPs are taken up by the nasal epithelium and the dendritic cells and macrophages in the nasal mucosa[125]. In an example using SARS-CoV-2 spike protein mRNA LNPs, the spike protein can be produced locally and drained to lymph nodes for presentation and induce immune response[125].

### 5.1.3 Enteral Route

The enteral route is medication delivery through oral ingestion into the gastrointestinal (GI) tract. Oral drug delivery is a common and convenient method, especially when delivering therapeutic agents targeting the GI tract. However, harsh environments in the GI tract, including steep pH change, pepsin, bile salts, mucin, mucosa, and the enterocytes barrier at the intestine, present significant challenges to delivering LNPs effectively[126]. Screening through different formulations can aid in the design of LNPs that are capable of oral delivery. However, existing work primarily focuses on LBNPs without ionizable lipid compounds, as the pH responsiveness can be a disadvantage when encountering complex pH conditions in the GI tract. One siRNA-LBNPs formed using $306O_{13}$ lipidoid[127], which has three fatty acid chains, was successfully taken up at the intestine and colon cells. Despite the successful intake, no significant gene silencing was observed[126]. Natural plant-based lipid is also explored in the context of oral delivery, providing higher stability in GI tract while potentially offering higher biocompatibility due to the nature of the lipid, with which several medicines treating ulcerative colitis (UC) are developed. LNPs can be formulated with lipids derived from ginger nanoparticles, which maintain stability in the GI tract and effectively target the colon. They can deliver mRNA to the target site and promote healing during UC[128]. Natural lipids extracted from mulberry leaves are also used in LNPs formulation, and significant gene silencing after delivery of LNPs with the CRISPER Cas9 system[129] was observed for UC. An enzyme-responsive system is also developed by utilizing esterase presence in the UC area, and drug release is triggered by the cleavage of chitosan by esterase at the target site[130].

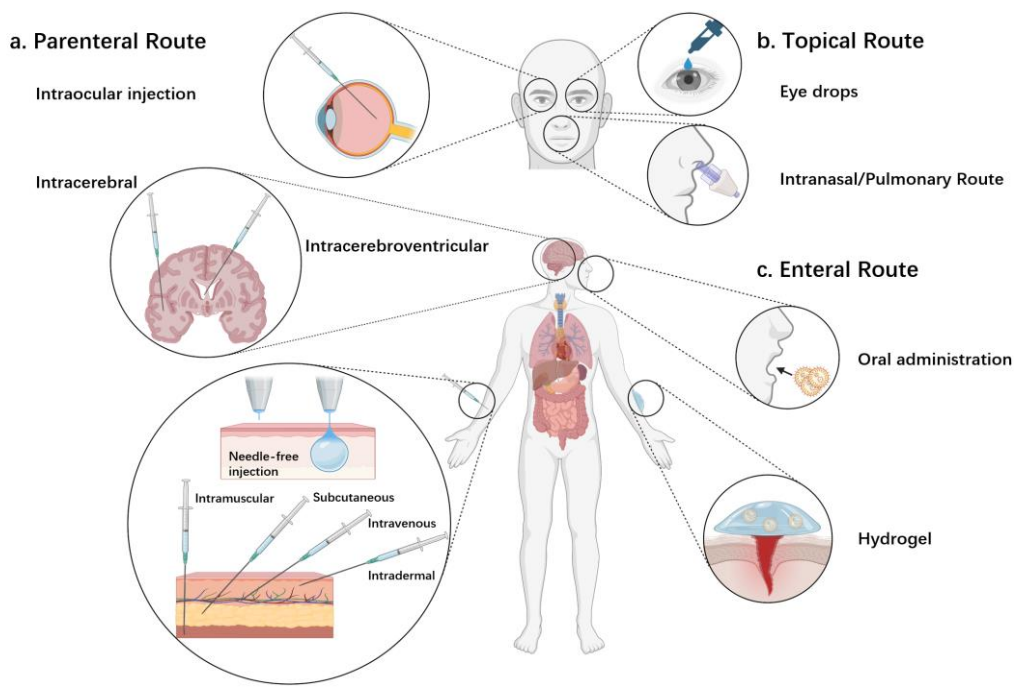

**Fig. 6 Dispersal Domain: Administration Route for LNPs. a.** Parenteral Route involves injection with a needle. **b.** Topical Route refers to applying drugs through the surface of the skin or mucosa. **c.** Enteral Route refers to oral intake in the context of LNPs.

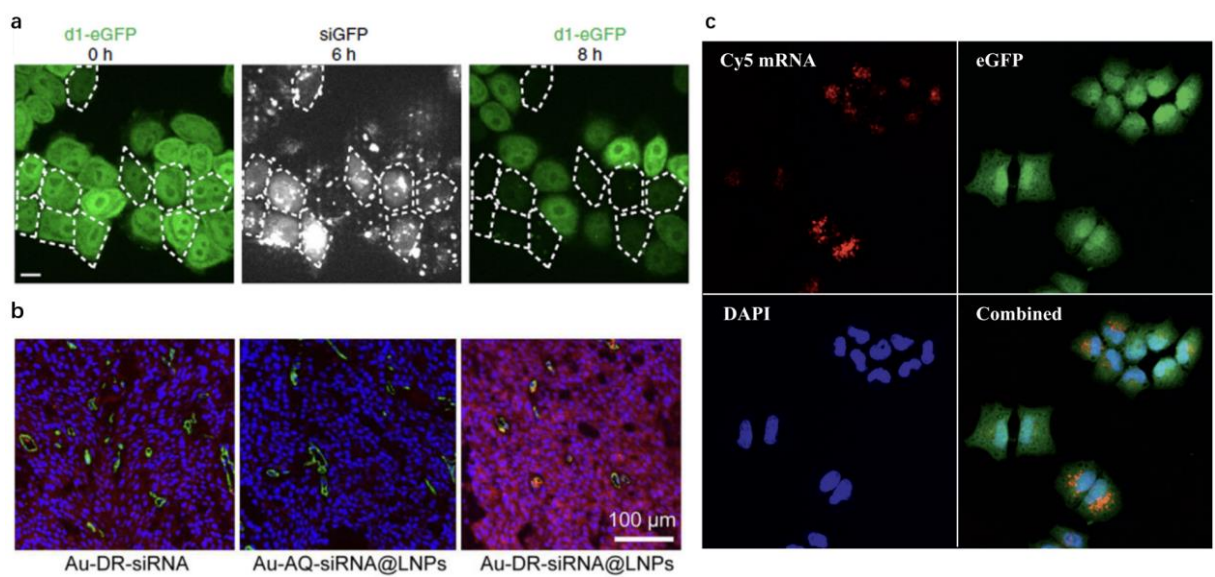

**Fig. 7 *In vivo* dynamics of LNPs. a.** Time-lapse fluorescence microscopy of HeLa-d1-eGFP cells treated with siGFP-Cy3 lipoplexes. Images show eGFP fluorescence at the start ($t = 0$ hours, left) and end ($t = 8$ hours, right) of the experiment. The middle panel outlines cells exhibiting cytosolic release events of siGFP-Cy3 within the first 6 hours, demonstrating the kinetics of siRNA delivery and target gene silencing. Scale bar: 10 µm. Adapted from ref[131] with permission. © 2024 Springer Nature Limited. **b.** Confocal laser scanning microscopy (CLSM) images of tumor tissue sections after treatment with LNP-encapsulated Cy5-labeled siRNA. Red fluorescence indicates the presence and localization of released siRNA within tumor cells. This visualization demonstrates the tumor-targeting capability of LNPs and the intracellular release of functional siRNA, which transitions from a quenched to an active state upon Dicer-mediated cleavage. Adapted from ref[104] with permission. © 2022 Shenyang Pharmaceutical University. Published by Elsevier B.V. **c.** Schematic representation of intercellular LNP-mRNA delivery via exosomes. Cells treated with LNPs containing Cy5-eGFP mRNA secrete exosomes containing the delivered mRNA, potentially facilitating its spread to neighboring cells. This mechanism suggests a novel mode of LNP-mediated nucleic acid dissemination beyond direct cellular uptake. Adapted from ref[132] with permission. © 1999-2024 John Wiley & Sons, Inc or related companies.

### 5.2 Intracellular Trajectory of LNPs: Navigating Endosomal Pathways and Cytosolic Delivery for Enhanced Therapeutic Efficacy

Once LNPs reach their target cells, their intracellular journey begins, marking a critical phase in the delivery of therapeutic payloads. The process of cellular uptake, typically via endocytosis, initiates a complex series of intracellular events that ultimately determine the fate and efficacy of the encapsulated drug. Recent studies have provided detailed insights into the kinetics of this uptake process. For instance, in HTB-177 cells, approximately 80% of radiolabeled LNPs disappeared from cell culture supernatants within 5 hours, indicating rapid cellular internalization[132]. Moreover, flow cytometry analysis revealed that over 90% of cells were positive for Cy5-mRNA within 2 hours of LNPs administration, demonstrating the efficiency of LNP-mediated mRNA delivery.

Upon internalization, LNPs are initially contained within early endosomes (EE). These vesicles represent the first sorting station in the endocytic pathway. While the mildly acidic environment (pH ~6.5) of early endosomes does trigger some degree of change in LNPs, the main stage for endosome escape occurs in the subsequent phase. As trafficking progresses, early endosomes mature into late endosomes (LE), characterized by a further drop in pH (to ~5.5) and the accumulation of specific protein markers. This increasingly acidic environment poses challenges for payload stability but also offers opportunities for triggered release mechanisms. The "proton sponge" effect, combined with the phase transition of LNPs mediated by ionizable lipids, becomes increasingly significant within late endosomes, enhancing endosomal escape, which has been visualized via fluorescence microscopy[131] (**Fig. 7a**).

The cytosol represents the ultimate destination for many LNPs payloads, particularly for nucleic acid therapeutics that exert their effects in this compartment. The dynamics of intracellular drug release from LNPs are complex and can vary depending on the specific payload and LNPs design. For nucleic acid cargoes, dissociating the payload from its lipid carrier in the cytosol is crucial for therapeutic activity. This process is influenced by factors such as the strength of electrostatic interactions between the payload and cationic lipids, the presence of cytosolic proteins that may compete for binding, and the stability of the payload itself. Recent research has provided quantitative insights into the efficiency of this process. In HTB-177 cells, it was observed that from $6.5 \times 10^{12}$ VEGF-A mRNA copies delivered via LNPs, approximately $1 \times 10^{12}$ VEGF-A protein copies were detected after 24 hours[132]. This translates to about 0.15 copies of VEGF-A protein produced per mRNA copy, highlighting both the potential and current limitations of LNP-mediated mRNA delivery and translation. The real-time dissociation of siRNA from LNPs in the cytosol was visualized by FRET techniques, providing valuable insights into the kinetics of this process and informing the design of more efficient delivery systems[104] (**Fig. 7b**). Interestingly, a fraction of internalized LNP-delivered mRNA can be packaged into exosomes and secreted, potentially extending the functional reach of the therapeutic mRNA beyond the initially targeted cells[41,132] (**Fig. 7c**). The secreted exosomes containing the therapeutic mRNA can then be taken up by other cells, as demonstrated by the delivery of Cy5-eGFP mRNA via exosomes to HTB-177 cells, resulting in eGFP protein production[132].

In conclusion, the dispersal domain of LNPs encompasses a complex journey from initial administration to intracellular payload delivery. Recent advances in LNPs design, imaging techniques, and mechanistic studies have significantly enhanced our understanding and control over LNPs behavior *in vivo*. The detailed kinetics of LNPs uptake, intracellular processing, and the newly discovered exosomes-mediated redistribution of mRNA add layers of complexity to this process. As we continue to unravel the intricacies of LNPs dispersal dynamics, we move closer to realizing the full potential of these versatile nanocarriers to achieve biology-level intelligence.

## 6 How are Artificial LNPs different from Natural Exosomes?

As we move toward more sophisticated drug delivery systems, it is crucial to recognize the parallels between artificial LNPs and natural exosomes. Here, we demonstrated the potential of our Four Domain Model—Architecture, Interface, Payload, and Dispersal—to provide a structured framework that not only elucidates the thermodynamic behavior of LNPs but also serves as a roadmap for future innovations.

Regarding the Architecture Domain, exosomes hold a lipid layer membrane, which is more similar to liposomes compared to LNPs, with size variation from 40 – 160 nm in diameter, comparable to LNPs[133]. Lipid composition differs from that of LNPs, emphasizing the inclusion of Phosphatidylserine (PtdSer)[134], which can be used for detection and diagnosis. For instance, PtdSer might be helpful for early cancer detection in mice[133].

As for the Interface Domain, unlike LNPs, surface proteins are usually observed in exosomes, and their roles include activating antitumor response[135], cell targeting[136], adhesion[137], and signaling[138], antigen presentation[139], protection from phagocytosis[140], and enzymes that modulate the biology process[141]. In contrast, the surface of LNPs, while potentially able to incorporate a broader category of substances, has much fewer kinds of bioreactivity compared to exosomes and mainly focuses on targeted release. While potentially safer, this also indicates that there is still much room for the programmable LNP intelligence to grow.

Another huge gap between LNPs and exosomes is in the Payload Domain. While current payloads in artificial LNPs demonstrate various therapeutic effects and exhibit programmable features (**Table 3**), extending the payload domain by examining natural exosome systems offers valuable insights for enhancing LNP versatility and efficacy. Protein is an important payload encapsulated in exosomes, while less is the case in LNPs. These proteins can be transcription factors[142] or histones[143] that regulate the gene expression and DNA packaging in the cell, which can be a promising approach adapted by LNPs. However, encapsulating proteins in LNPs is not as straightforward as encapsulating negatively charged nucleic acids, and advances in this field are required. Metabolites small molecules such as amino acids, nucleotides, and citric cycle intermediates also exist in exosomes, potentially affecting metabolism in the recipient cell[133]. Encapsulation of small molecules in LNPs, while it can lead to promising applications, can be challenging as LNPs can act as semi-closed systems and have difficulty keeping small molecules from diffusing without additional molecular-level forces.

The Dispersal Domains for the two systems are also different due to their different origin and whether they are required to be administered. However, as they mainly function in biological environments, comparing these two systems is reasonable and will give important insights into intelligent systems in the body. In conclusion, LNPs are different from exosomes in all four domains, demonstrating the effectiveness of understanding LNPs and similar systems. These insightful comparisons are expected to significantly advance the field of programmable drug delivery systems.

## 7 Ethical Considerations

The development of LNPs has prompted ethical considerations that must be addressed to ensure responsible innovation and clinical application. A primary concern is the safety of these systems, which can be affected by the complex and potentially hidden side effects of lipid and nucleic acid components, as highlighted by the SARS-CoV-2 mRNA vaccines. Additionally, the evolving understanding of in vivo effects of nucleic acids and other therapeutic agents, especially in the context of sophisticated programmable LNPs, requires a balance between development speed and safety.

Predictability is another major concern as programmable LNPs integrate diverse functional components and complex therapeutic mechanisms. This increases the complexity of interaction with biological environments, necessitating a thorough understanding of physiology to predict pharmacodynamics and pharmacokinetics, thereby minimizing risks of unexpected side effects and long-term health issues.

As programmable LNPs approach the complexity of living systems, ethical considerations relating to artificial life emerge. If LNPs can navigate the body and provide therapeutic effects, they may be perceived as artificial exosomes or viruses. Their potential environmental impact also arises, as these complex systems could interact with natural cells, bacteria, and viruses, requiring rigorous assessment criteria beyond clinical considerations.

Ultimately, ethical questions arise regarding the accessibility of sophisticated programmable LNPs. As their complexity increases, so does the likelihood of higher costs and the need for specialized expertise and equipment, potentially exacerbating healthcare inequality. Ensuring equitable access to these technologies is a significant challenge.

## 8 Conclusion and Outlook

In conclusion, this research presents a new conceptual framework that breaks down the key components of programmable LNPs into four distinct domains: Architecture, Interface, Payload, and Dispersal Domain. This modular model, akin to Lego blocks, enables the rational design of programable LNPs for precise payload control and reduced off-target effects. Unlike traditional models such as liposome models[144] and polymeric nanoparticle models[145], which often treat LNPs as an indivisible entity, the Four-Domain Model identifies LNPs LNPs as dynamic, semi-closed systems in equilibrium, acknowledging potential matter exchange due to their porous surfaces and diffusion process[40]. **Table 4** provides a comprehensive overview of various stimuli-responsive mechanisms employed in programmable LNPs across these domains. We hope this novel model can accelerate advancements in precision medicine and programmable drug delivery systems.

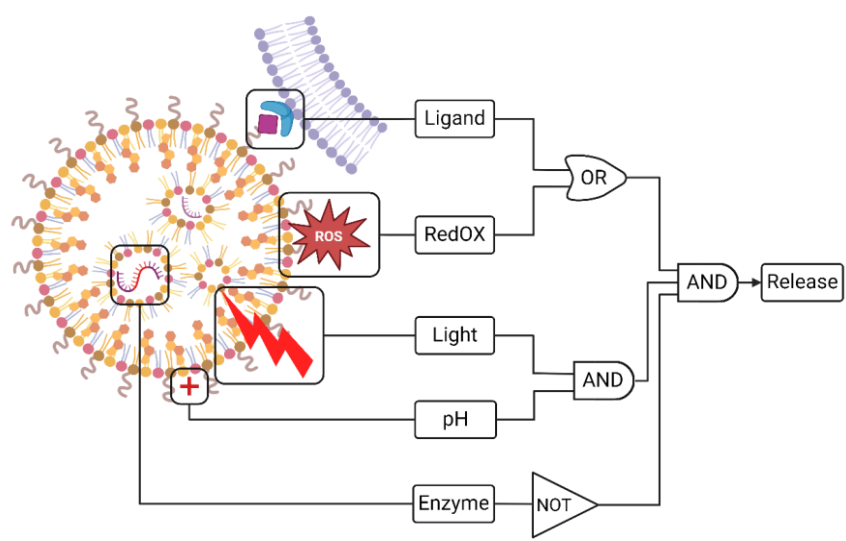

**Fig. 8 Outlook of programmable LNPs.** The future of programmable LNPs involves integrating multiple stimuli-responsive elements to achieve precise and controlled therapeutic cargo release under specific conditions, as illustrated in this schematic. AND gate indicates release or therapeutic activity only when both stimuli are present. OR gate indicates release when only one of the stimuli is present. The NOT gate indicates the failure of the release of loss of therapeutic effect when a certain input stimulus is present.

Despite notable progress in the development of programmable LNPs, liposomes remain dominant in the therapeutic environment as seen by their established track record (**Table 5**). Programmable LNPs must overcome development and clinical translation obstacles in order to go from theoretical promise to real-world use. This involves refining their design, ensuring regulatory compliance, optimizing clinical trials, engaging with stakeholders, and establishing scalable manufacturing processes. Post-approval collection of real-world evidence further substantiates their long-term safety and efficacy, facilitating broader clinical adoption.

The inherent complexity of programmable LNPs, due to their multi-component nature, poses challenges in design and formulation. The vast number of variables in LNP design makes exhaustive testing of all potential formulations difficult and inefficient, risking the omission of optimal combinations. Recent developments in machine learning have introduced innovative methods that enable the prediction of therapeutic outcomes by analyzing the properties of individual components. Machine learning advancements, such as the AGILE platform using deep learning with graph neural networks and molecular descriptors[146], and the TransLNPs model based on transformers[147], have shown promise in predicting mRNA transfection potency and efficiency. These tools streamline LNP design, enabling more efficient exploration of the high-dimensional design space and potentially uncovering novel delivery mechanisms. On the other hand, the current microfluidic formulation techniques might not be universally applicable due to the complex physiochemical properties of novel functional components. Engineering techniques for proteins to achieve orthogonal modification are developed to face this challenge. However, to truly unleash the potential of programmable LNPs, a novel universal engineering technique is required.

In addition to the complexity of refining programmable LNP designs, clinical adaptation presents another challenge. Novel materials contain a variety of functional aspects, thus, it's important to assess their metabolism, bioactivity, biocompatibility, and potential side effects. The intricacy of these combined impacts might be more than current clinical trials can handle. Therefore, Clinical trial protocols must be updated to consider the complexity of these cutting-edge technologies.

This review compared artificial LNPs and natural exosomes with our Four Domain Model, highlighting their functionality, biocompatibility, and therapeutic potential. Programmable LNPs provide precise control over therapeutic payloads and can be engineered for targeted delivery, while natural exosomes offer a biocompatible, endogenous platform with the potential for reduced immunogenicity. LNPs benefit from scalable manufacturing and real-time monitoring capabilities but face design complexity and regulatory challenges. With their natural cargo and targeting properties, exosomes struggle with production scalability and consistency. The choice between these delivery systems depends on the therapeutic application, balancing the need for customization and control against biocompatibility and natural delivery mechanisms. This comparative analysis also provides a roadmap for developing more intelligent, naturally integrated systems.

Nevertheless, the existing artificial interface remains simply a "boundary" of a thermodynamic system instead of an intelligent gatekeeper for complex molecular fluxes, as provisioned in **Fig. 8**. As programmable LNPs evolve to incorporate multiple functions and stages, the outlined framework serves as a solid basis for overcoming challenges and maximizing their potential in various therapeutic contexts. Furthermore, we discussed novel anatomical delivery routes, such as intranasal and intraocular methods, that are associated with their challenges and opportunities. We prospect for future research directions to maximize LNP potential in precision medicine for the ultimate clinical application. When producing a well-controlled and adaptative artificial interface becomes a reality, so will artificial functional vesicles.

## 10  Acknowledgments

The authors are grateful for the funding support from the Hong Kong Research Grants Council (project reference: GRF14204621, GRF14207920, and GRF14207419).

The authors would like to acknowledge Ms. Yu Xiao and Ms. Syeda Aimen Abbasi (Department of Biomedical Engineering, The Chinese University of Hong Kong) for their support in the project development. The authors would like to thank BioRender.com for providing the tools used to create the illustrations in this article. The authors would like to extend their gratitude to GPT-4o for providing the tools that significantly enhanced the clarity and quality of the content in this article.

## 11  Author information

Authors and Affiliations

**Department of Biomedical Engineering, The Chinese University of Hong Kong, Shatin, Hong Kong SAR, 999077, China**

Zhaoyu Liu, Jingxun Chen, Yuanyuan Wei and Ho-Pui Ho

**Department of Biomedical Engineering, Whiting School of Engineering, Johns Hopkins University, Baltimore, Maryland, 21218, USA**

Zhaoyu Liu, Jingxun Chen

**Guangdong Institute of Intelligence Science and Technology, Hengqin, Zhuhai, 519031, China.**

Mingkun Xu

**Department of Chemical and Biomolecular Engineering, Johns Hopkins University, Baltimore, Maryland, 21218, USA.**

David H. Gracias

**School of Biomedical Engineering, The University of Sydney, Sydney, New South Wales 2006, Australia.**

Ken-Tye Yong

**The Biophotonics and Mechano-Bioengineering Lab, The University of Sydney, Sydney, New South Wales 2006, Australia.**

Ken-Tye Yong

**Department of Neurology, David Geffen School of Medicine, University of California, Los Angeles, California, 90095, USA**

Yuanyuan Wei

Contributions

     Z. Liu and Y. Wei conceived and designed the review theme; Z. Liu gathered and analyzed the literature; Z. Liu and Y. Wei drafted the manuscript; Z. Liu created the tables; J. Chen created the figures; Y. Wei, H. Ho, K. Yong, M. Xu, and D. Gracias critically revised the manuscript for important intellectual content; H. Ho oversaw the project and acquired funding. All authors discussed the results and implications and commented on the manuscript at all stages.

Corresponding author

Correspondence to Ken-Tye Yong, Yuanyuan Wei, and Ho-Pui Ho.

## 12   Ethics declarations

No conflict of interest

**Additional information**

Table 1. Comparative analysis of formulation methods for programmable LNPs.

| Method | Description | Drug degradation | Scalability | Size & PDI Control | Advantages | Encapsulation Efficiency | Disadvantages | Complexity | Accessible Volume | Applicability to Different Payloads | Reference |
|---|---|---|---|---|---|---|---|---|---|---|---|
| **Microfluidics Approaches** | Introduction of turbulence in a microfluidic channel facilitates the formation of LNPs. | Low | High | Excellent | High reproducibility | Excellent | Cost of initial investment and possibility of clogging in microchannels | Moderate | uL to L | Broad (small molecules, proteins, nucleic acids) | Ref. [50,148] |
| **Ethanol Injection** | Dissolution of the lipid matrix in a water-miscible solvent followed by rapid injection into a mixed aqueous phase under constant stirring. | Low | Medium | Average | Versatile and of simple implementation | Good | Residual solvent, heterogenous | Low | uL to mL | Limited (small molecules, some proteins) | Ref. [148,149] |
| **Thin-film Hydration** | Dissolution of lipids in organic solvents followed by evaporation forms a thin film that, upon hydration in aqueous media, produces LNPs. | Medium | Medium | Poor | Cost effective and simple | Low | Extra steps needed for homogeneity (i.e., extrusion), time consuming | Low | uL to mL | Broad (small molecules, nucleic acids) | Ref. [148,150] |
| **T-junction Mixing** | T-Junction mixing involves using high-speed impingement jet mixers for flash nanoprecipitation, where solvent and anti-solvent phases are mixed nearly instantly, facilitating the efficient production of nanoparticles. | Medium | High | Excellent | Efficient and high throughput | Good | Requires large volumes of liquids, making it less suitable for small-scale or early-stage development. | Moderate | mL to L | Limited (small molecules, some proteins) | Ref. [148] |
| **Microemulsion** | Microemulsions create LNPs by forming nanometer-sized droplets where lipids self-assemble at the interface, stabilized by surfactants, allowing controlled nucleation and growth. | Medium | Low | Poor | Suitable for small volume vs. surfactants | Average | High proportion of surfactants (20-40%) | Low | uL to mL | Limited (small molecules, some proteins) | Ref. [148,151] |
| **Microfluidic Vortex Mixing** | Utilization of a vortex flow field to achieve high throughput synthesis of size-tunable lipid vesicles by focusing the lipid stream within a rotating aqueous buffer. | Low | High | Good | Efficient and high throughput | Good | May need downstream processing for concentration and purification. | Moderate | mL | Broad (small molecules, nucleic acids) | Ref. [152] |
| **Spray Drying** | Atomization of a liquid feed containing dissolved or suspended LNPs into a hot drying gas stream to produce a dry powder. | Medium (7.5-14% siRNA loss) | High | Good | Produces dry powder suitable for inhalation and Single-step process | Good | Heat exposure can damage thermolabile components | Moderate | mL to L | Broad (small molecules, nucleic acids) | Ref. [153,154] |

Table 2. Overview of computational models for LNPs design and analysis.

| Computational Model | Model Principle | Application | Strength | Weakness | Typical Software/Tools | Time Scale | Data Input Requirements | Reference |
|---|---|---|---|---|---|---|---|---|
| All Atom Molecular Dynamics | Simulates atomic interactions over time using classical mechanics. | Study LNPs formation, structure, and RNA encapsulation; Examine effects of pH, lipid composition, and RNA on LNPs properties; Investigate lipid phase behavior and RNA-lipid interactions. | Provides atomistic insights into LNPs structure and dynamics. Can simulate self-assembly process. Allows systematic variation of LNPs composition and conditions. | Limited to small system sizes and short timescales compared to real LNPs. Force field accuracy can impact results. Difficult to directly compare to experimental LNPs characterization. | GROMACS, NAMD, OpenMM, AMBER | Nanoseconds to microseconds | Lipid and RNA structures and parameters; System composition and initial configuration; Simulation parameters (temperature, pressure, etc.). | Ref. [155–157] |
| Coarse-Grained Molecular Dynamics Simulations | Reduces atomic detail by grouping atoms into larger "beads". | Study LNPs self-assembly and structure. Investigate LNP-membrane fusion and RNA release. Examine effects of LNPs size and composition. | Enables simulation of larger systems over longer time scales; computationally less demanding than full MD. Allows study of collective lipid behavior and phase transitions. | Loses atomic level detail. Force field development and validation is challenging. Mapping between CG and atomistic representations not always straightforward. | MARTINI, SIRAH, ELBA, SDK | Microseconds to milliseconds | CG mapping scheme for lipids and RNA; CG force field parameters; System composition and initial configuration. | Ref. [158–160] |
| Monte Carlo Simulations | Uses probabilistic methods to explore configurations; can handle a vast parameter space. For LNPs, they often use coarse-grained models and can access larger time and length scales than all-atom MD. | Study pH-dependent behavior of LNPs. Investigate lipid ionization states. Examine lipid packing and phase behavior. Explore RNA-lipid interactions in LNPs core. | Effective for sampling large configuration spaces; can incorporate complex energy landscapes; can access larger system sizes and longer timescales. | May require extensive sampling to achieve convergence; lacks explicit dynamics information; may not capture kinetic effects. | LAMMPS, HOOMD-blue | Varies widely | Coarse-grained models for lipids and RNA; system composition; Interaction potentials; Simulation parameters (temperature, chemical potentials, etc.). | Ref. [160,161] |
| Poisson-Boltzmann Calculations | Solves the Poisson-Boltzmann equation to model electrostatic interactions in solution. | Predict electrostatic potential distributions in LNPs; Estimate lipid ionization states; Study pH-dependent behavior of LNPs core; Investigate ionic strength effects on LNPs structure. | Computationally efficient compared to molecular simulations; Can handle large systems and long-range electrostatic effects; Provides continuum description of ionic atmospheres; Allows systematic variation of solution conditions (pH, ionic strength). | Neglects ion-ion correlations and ion-specific effects; May not capture molecular-level details or lipid packing effect; Assumes thermal equilibrium; Simplified geometry may not fully represent complex LNPs structures. | APBS (Adaptive Poisson-Boltzmann Solver), DelPhi, MEAD | Not applicable - provides equilibrium electrostatic properties | System geometry (e.g., cylindrical model of LNPs core); Charge distributions or fixed charges; Dielectric constants for different regions; Ionic strengths and species; pH and pKa values for ionizable groups. | Ref. [160,162] |
| PBPK/PD Model for LNP-mediated mRNA Delivery | Physiologically based pharmacokinetic/ pharmacodynamic modeling that accounts for LNPs distribution, degradation, and mRNA expression. | Predict biodistribution and transgene expression of mRNA-LNPs in mice; Evaluate effects of targeting on LNPs behavior. | Bridges gap between LNPs delivery kinetics and delayed transgene expression; Accounts for physiological factors and active targeting; Can be expanded to various conditions and species. | Requires experimental data for parameter estimation and validation; May not capture all molecular-level interactions. | ADAPT 5 (BMSR) | Hours to days | Blood and tissue PK data; Luciferase expression kinetics; Physiological parameters (organ volumes, blood flows); LNPs properties. | Ref. [103] |
| Machine Learning Based | Utilizes conventional machine learning models such as support vector machines (SVM), random forests, and other regression techniques to predict key LNPs characteristics. | Predicts LNPs properties like encapsulation efficiency, transfection efficiency, and stability by leveraging structural and physicochemical features of LNPs components. | Offers rapid and scalable predictions across diverse datasets; Efficiently captures nonlinear relationships in data; Easily integrates with high-throughput screening workflows; Suitable for exploratory data analysis and feature selection. | Performance may be limited when handling complex systems with intricate interactions; Model accuracy heavily depends on the quality and diversity of training data; Generalization to novel lipid structures can be challenging. | Scipy, MATLAB | Not applicable – provides predictions of static properties rather than dynamic behavior. | LNPs characteristics, including molecular descriptors, structural features (e.g., SMILES), and experimental measurements (e.g., encapsulation efficiency, transfection efficiency). | Ref. [103,146,147] |
| AGILE | The AGILE platform uses a deep learning approach combining graph neural networks (GNNs) and molecular descriptor encoding to predict the mRNA transfection potency of ionizable lipids. | Accelerating the discovery and optimization of LNPs for mRNA delivery to specific cell types. | Rapid screening of large virtual lipid libraries Ability to identify novel lipid structures. Cell-type specific optimization Integration of structural and molecular property data. Interpretable results providing insights into structure-activity relationships | Reliance on experimental data for fine-tuning, which can be time-consuming to generate. Potential limitations in predicting performance of lipid structures significantly different from the training set. | PyTorch, RDKit | Not applicable – provides design but rather kinetics or dynamics | SMILES representations of lipid structures; Molecular descriptors calculated using Mordred.; Experimental mRNA transfection potency data for model fine-tuning. | Ref. [146] |
| **TransLNP** | TransLNPs is a transformer-based deep learning model that combines pretraining on large molecular datasets with fine-tuning on limited LNPs data to predict the transfection efficiency of ionizable LNPs for mRNA delivery. | Accelerating the screening and optimization of LNPs for mRNA delivery by predicting their transfection efficiency. | Combines both coarse-grained atomic sequence and fine-grained 3D structural information. Utilizes pretraining on large molecular datasets to improve performance on limited LNPs data. Incorporates data balancing techniques (BalMol block) to manage imbalanced datasets Outperforms AGILE in predicting LNPs transfection efficiency. | Still requires experimental data for fine-tuning, which can be time-consuming to generate. Performance may be limited by the "transfection cliff" phenomenon, where small structural changes lead to significant differences in transfection efficiency. | PyTorch, RDKit, Custom BalMol block for data balancing | Not applicable – provides design but rather kinetics or dynamics | SMILES representations of lipid structures; 3D conformational information for molecules; Experimental transfection efficiency data for fine-tuning. | Ref. [147] |

**Table 3. Characteristics and applications of payloads encapsulated in programmable LNPs[1].**

| Payload Type | Example | Clinical Applications | Solubility | Charge | Encapsulation Location | Encapsulation Efficiency | Stability Consideration | | Stability Concerns & Disadvantages of Direct Delivery | Packaging and Delivery Enhancements | Half-life in Circulation | Mechanism of Action | Reference |
|---|---|---|---|---|---|---|---|---|---|---|---|---|---|
| | | | | | | | Storage Conditions | Formulation Additives | | | | | |
| mRNA | mRNA vaccines, eGFP mRNA, Luciferase mRNA, IL-2 mRNA | Gene therapy, vaccines, reporter gene expression, immunotherapy | Hydrophilic | Negative | Aqueous core | 82-95% | -80°C to -20°C | Lipids, PEGylated lipids, cholesterol | Rapid degradation by RNases, poor cellular uptake, immune stimulation | Lipid formulation enhances delivery | Minutes to hours | Induces protein expression | Ref. [13,90] |
| siRNA | PKN3 shRNA | Gene silencing | Hydrophilic | Negative | Aqueous core | >93% | -80°C to -20°C | Lipids, PEGylated lipids, cholesterol | Rapid degradation by RNases, poor cellular uptake, off-target effects | Targeted delivery to specific tissues | Minutes to hours | RNA interference (RNAi) | Ref. [93,94] |
| microRNA | miR-193b-3p mimic, miR-21, miR-18a, let-7c | Gene therapy, Cancer therapy | Hydrophilic | Negative | Aqueous core | >90% | -80°C to -20°C | Lipids, PEGylated lipids, cholesterol | Rapid degradation by RNases, Poor cellular uptake | Enhanced delivery with LNPs | Minutes to hours | Regulation of gene expression, Modulation of translation | Ref. [96–98] |
| Self-amplifying RNA | SARS-CoV-2 spike protein RBD, influenza hemagglutinin | Vaccines | Hydrophilic | Negative | Aqueous core | >90% | -80°C to -20°C | Lipids, PEGylated lipids, cholesterol | Rapid degradation by RNases, poor cellular uptake, potential for immunogenicity | Lipid formulation enhances delivery | Hours to days (longer than conventional mRNA) | Self-amplification in cytoplasm leads to elevated levels of antigen expression, inducing strong immune responses | Ref. [91,92] |
| Translation-activating RNA | PTV-IIIab-based taRNA, mini taRNA (94 nt) | Potential for treating haploinsufficiency diseases | Hydrophilic | Negative | Aqueous core | - | - | Formulation Additives: Stabilizing hairpins added to 5' and 3' ends | Transient effect (12-24 hours) when delivered as unmodified RNA | AAV delivery also demonstrated | - | Recruiting translation initiation factors to target mRNA to enhance translation | Ref. [163] |
| CRISPR Components | Cas9 mRNA, sgRNA | Gene editing | Hydrophilic | Negative | Aqueous core | >90% for most formulations | -80°C to -20°C | Stabilizers, enzymes | Rapid degradation, potential off-target effects, immune responses | Enhanced delivery with LNPs | Minutes to hours | Genome editing | Ref. [99,100] |
| DNA | Plasmid DNA (pDNA) | Gene therapy | Hydrophilic | Negative | Aqueous core | ~60-80% | -80°C to -20°C | Lipids, polyethyleneimine | Poor cellular uptake, nuclear entry challenges, immune responses | Enhanced cellular uptake | Hours to days | Gene expression | Ref. [20,93] |
| Antifungals | Fluconazole (Fluc) | Antifungal therapy | Hydrophilic | Neutral, subject to ionization | Aqueous core, lipid bilayer | ~59% | - | - | Poor aqueous solubility, limited bioavailability | Enhanced solubility and stability | - | Inhibition of fungal growth | Ref. [40] |
| **Chemotherapeutics** | Dox, Paclitaxel | Cancer therapy | Amphiphilic (Dox), Hydrophobic (Paclitaxel) | Neutral (Paclitaxel), Positive (Dox) | Aqueous core & lipid bilayer (Dox), Lipid bilayer (Paclitaxel) | ~100% (Dox), ~95% (Paclitaxel) | - | - | Cardiotoxicity, myelosuppression (Dox), Poor aqueous solubility, systemic toxicity (Paclitaxel) | Targeted delivery reduces side effects | - | Cytotoxicity | Ref. [101,102] |
| **Enzymes** | Proenzymes, Coenzyme Q10 | Enzyme replacement therapy, metabolic disorders | Hydrophilic/hydrophobic (CoQ10) | Variable | Aqueous core/lipid layer (CoQ10) | ~70% | -20°C to -80°C | Stabilizers, cryoprotectants | Rapid clearance, immunogenicity, denaturation | Improves stability and activity | - | Enzyme replacement | Ref. [22,164] |
| **Metal Nanoparticles** | Gold nanoparticles | Photothermal therapy | Generally hydrophobic | Neutral to negative | Aqueous core, surface modified | 92 ± 5% | - | - | Potential toxicity, aggregation, stability under physiological conditions | - | - | Photothermal effect for targeted drug release | Ref. [84] |

Table 4. Overview of stimuli-responsive mechanisms across programmable LNPs domains.

| Domain | Input | Functional Component | FDA Approval Status | Mechanism | Characterization Techniques | Application | Functional Component Synthesis | LNPs Formation | Encapsulation Efficiency | Payload | Reference |
|---|---|---|---|---|---|---|---|---|---|---|---|
| Architecture | pH | DLin-MC3-DMA (MC3) | Approved as part of mRNA-based vaccines (COVID-19) | transition from inverse spherical micellar to inversehexagonal lipid structures in lipid excess phases. | SAXS, Cryo-TEM, DLS, Fluorescence spectroscopy, Gradient UPLC, bDNA analysis, ELISA, Fluorescence microscopy, Fluorescence anisotropy | mRNA Delivery | Purchased from AstraZeneca | ethanol injection method | Not specified | mRNA | Ref. [39] |
| | | Phytantriol-based Ionizable Lipid Nanoparticles (ALC-0315/PT, SM-102/PT) | Components like ALC-0315 and SM-102 approved in COVID-19 mRNA vaccines | Lyotropic liquid crystalline mesophase transitions from cubic to hexagonal structures with pH changes to enhance endosomal escape. | DLS, SAXS, 2-(p-toluidino) naphthalene-6-sulfonic acid assay, Zeta Potential Measurement | Drug Delivery | Purchased from Advanced Molecular Technologies (Scoresby, Victoria, Australia) | thin-film hydration method followed by probe sonication | Not specified | Various | Ref. [33] |
| | | 2-morpholinoethyl oleate (O2ME) | No specific approval; research stage | Transition from hexagonal (H2) to cubic (cubosomes) phase with pH changes, which interacts with Fungal cells. | NMR, DLS, SAXS, HPLC, UV-vis, CLSM, SEM | Antifungal Therapy | Esterification reaction between oleic acid and 2-(2-hydroxyethyl) morpholine using EDC and DMAP as catalysts, followed by purification using flash chromatography | thin-film hydration method followed by probe sonication | 59 | Fluconazole (Fluc) | Ref. [40] |
| | Light | NIR-II dyes conjugated lipid (Cy-lipid) | No specific approval; research stage | Protonation of Cy-lipid in acidic conditions increases NIR-II absorption, facilitating photothermal endosomal escape. | NMR, HRMS, TEM, DLS, UV-vis-NIR, CLSM, Fluorescence spectroscopy, Thermal imaging, Fluorescence microscopy, Bioluminescence imaging | mRNA Delivery | Nucleophilic substitution between thiol lipid and polymethine dye | Mixing of lipids and mRNA in citrate buffer, followed by dialysis | ~90% | eGFP and luciferase mRNA | Ref. [83] |
| | | Porphyrin-lipid | No specific approval; research stage | Generates ROS upon NIR light irradiation, disrupting endosomal membranes. | DLS, TEM, cryo-EM, UV-vis, STED, RT-qPCR, Fluorescence spectroscopy, Confocal microscopy, Bioluminescence imaging | Gene | Acylation reaction between lysophosphatidylcholine and pyropheophorbide (a chlorophyll-derived porphyrin analogue) | Microfluidic rapid mixing | >95% | siRNA | Ref. [84] |
| Payload | Enzyme | IL-2F (IL-2, cleavable linker, CD25) | Components (like IL-2) approved, but not specific construct | Cleavage of linker by MMP-14 in tumor microenvironment, releasing active IL-2. | DLS, TEM, ELISA, CLSM, Immunohistochemistry, Western blot, Flow cytometry, Immunofluorescence, Hematoxylin and eosin (H&E) staining, TUNEL staining | Cancer | Cleavable linker (SGRSEN↓IRTA) inserted between IL-2 and CD25 sequences using PCR and Gibson assembly | Microfluidics with ethanol and aqueous phases | Not specified | IL-2 mRNA | Ref. [64] |
| | | NQO1-responsive proenzyme | No specific approval; research stage | Activation in the presence of overexpressed NQO1 in tumor cells. | MALDI-TOF MS, SDS-PAGE, TEM, DLS, T7EI assay | Cancer | The QPN ligand was attached to lysine residues at or near the active sites of the target proteins | Rapid mixing | Not specified | Proenzymes | Ref. [22] |
| | Light | 5 nm gold nanoparticles | Approved for some medical uses, not specifically in lipids | Localized heating of AuNPs causes thermal decomposition of adjacent lipid lamellae, releasing payload. | DLS, UV-vis, Cryo-TEM, FM, fluorescence spectroscopy | Cancer | AuNPs: Commercially obtained (Ted Pella Inc, 5 nm PELCO NanoXact Gold Nanoparticles) | T-tube protocol mixing lipids (DODAP, DSPC, cholesterol, PEG-DSPE) with AuNPs, followed by dialysis | ~100% | Doxorubicin | Ref. [106] |
| **Interface** | pH | imidazole based lipid (O12-D3-I3) | Not specifically approved; similar lipids used in approved formulations | Buffering induced Swelling enhance Endosomal Escape. | TEM, TNS fluorescence titration, DLS, colorimetric method, HPLC-MS | Cancer | Multi-step synthesis involving tert-butyl carbamate protection, amine alkylation, and imidazole derivatization | preformed vesicle method | 93 | siRNA | Ref. [76] |
| | | PGlu(DET-Car)30 LNPs | No specific approval; research stage | Active charged surface to interact with negative charged membrane. | DLS, TEM, Fluorescence Spectroscopy, FCM, CLSM, Bioluminescence Imaging, NMR Spectroscopy, HPLC | Cancer | Multi-step reaction involving biphasic solvent system and purification by acetonitrile precipitation and centrifugation | t-BuOH dilution procedure | 95 ± 3% | siRNA | Ref. [21] |
| | | double pH-responsive lipo-amino fatty acid (LAF) – Stp oligoaminoamide (OAA) carriers | No specific approval; research stage | Increased solubility and amphiphilic if carriers disrupt the membrane. | DLS, ELS, Agarose Gel Shift Assay, RiboGreen Assay, Flow Cytometry, MTT Assay, CLSM, RT-PCR, Gel Electrophoresis, In Vitro Cleavage Assay, PCR, Sanger Sequencing, TIDE Analysis, Ex Vivo Luciferase Assay | Gene Editing | use of Fmoc solid-phase assisted peptide synthesis (SPPS) to covalently connect Stp and LAF units via branching lysine, forming various topologies (combs, bundles, T-shapes, and U-shapes) with different Stp/LAF ratios. | microfluidic mixing method | >90% for most formulations | Cas9 mRNA/sgRNA | Ref. [165] |
| | | pH-activated lipid-like material (ssPalm) with vitamin E scaffolds | No specific approval; research stage | Destabilizing membranes at low pH by Tertiary amines. | DLS, PicoGreen, IVIS, FC, CLSM, ELISA | Vaccine | Performed mesylation and amination followed by acylation, stirred in chloroform with catalysts at room temperature. Purified by silica gel column chromatography using a chloroform | ethanol injection method | 94±4 | DNA | Ref. [20] |
| | Enzyme | Esterase-labile Quaternium Lipidoid (AMP-POC18) | No specific approval; research stage | Stable lipidoid underwent decationization to interact with endosome membrane. | DLS, Cryo-EM, HPLC, DSC, FACS, CLSM, IVIS, ELISA, ELISpot | mRNA Delivery | Combinatorial synthesis of quaternium lipidoids involving the reaction of amines with acrylate ester linkers followed by alkylation with bromides | Ethanol injection method, dialysis | 82 | mRNA | Ref. [82] |
| | | Phospholipid D (PLD) | Not specifically approved; general class used in various formulations | PLD enzymes cleave choline headgroups from phospholipids, forming membrane-bound anionic phosphatidic acid. | SPARTA, SAXS, SANS, Cryo-TEM, DLS, Fluorescence assays | neurodegeneration, vascular disease, and cancer | Purchased from Sigma Aldrich, used directly. | Thin-film hydration method followed by probe sonication | Not Provided | Not Provided | Ref. [23] |
| | | Polyphosphate (TPP) | Not specifically approved; used in research | Charge reversal triggered by ALP, enhancing cellular uptake and transfection efficiency. | DLS, ELS, FT-IR, UV-vis, Fluorescence spectroscopy, Agarose gel electrophoresis | Cancer | Polyphosphate coated on LNPs with Prot-Palm | Solvent injection method, followed by TPP coating | Not specified | pDNA | Ref. [64] |
| | Redox potential | Disulfide bonds | Common in biologics; specific use case approvals | GSH cleaves disulfide bonds within the hydrophobic tails of the lipids, leading to truncation and disintegration. | MALDI-TOF MS, SDS-PAGE, TEM, DLS, T7EI assay | Cancer | Michael Addition Reaction | Rapid mixing | Not specified | Proenzymes | Ref. [22] |
| | | BAmP-TK-12 (ionizable lipid with thioketal linkage) | No specific approval; research stage | Oxidation and cleavage of thioketal linkage by elevated ROS in tumor cells, destabilizing LNPs. | MALDI-TOF MS, SDS-PAGE, TEM, DLS, T7EI assay | Cancer | Not mentioned | Rapid mixing | Not specified | Proenzymes | Ref. [22] |

| | | Disulfide bonds in DDA-SS-DMA ionizable lipid | No specific approval; research stage | High GSH levels in tumor environment cleave disulfide bonds, triggering nanoparticle disassembly and payload release. | NMR, TOF-MS, DLS, TEM, CLSM, qRT-PCR, Agarose Gel Electrophoresis, Western Blot, Hematoxylin and Eosin (H&E) Staining | Cancer | 1) Oxidation of thioglycerol, 2) Esterification with dodecanoic acid, 3) Reaction with 2-(dimethylamino)ethane-1-thiol | Microfluidic rapid mixing | Not specified | PKN3 shRNA | Ref. 79 |
|---|---|---|---|---|---|---|---|---|---|---|---|

Table 5. Overview of clinical advancements in programmable lipid-based nanoparticles.

| Formulation/Brand Name | Targeting Mechanism | Product Overview | Target Disease/Clinical Trial | Mechanism of Action | Formulation Type | Administration Route | Clinical Trial Phase | Notable Results | Side Effects/Adverse Reactions | Market Availability | Clinical Trial Identifier |
|---|---|---|---|---|---|---|---|---|---|---|---|
| **Onivyde (Irinotecan Liposome)** | Enhanced permeability and retention (EPR) effect | Liposomal formulation of irinotecan | Metastatic pancreatic cancer | Inhibition of topoisomerase I | Liposome | Intravenous | Approved | Improved overall survival | Neutropenia, diarrhea | Available | - |
| **Doxil (Liposomal Doxorubicin)** | EPR effect | PEGylated liposomal doxorubicin | Multiple cancer types (Ovarian cancer, multiple myeloma, Kaposi's sarcoma) | Intercalates DNA, inhibits topoisomerase II | Liposome | Intravenous | Approved | Reduced cardiotoxicity compared to free doxorubicin | Hand-foot syndrome, mucositis | Available | - |
| **Myocet (Liposomal Doxorubicin)** | EPR effect | Non-PEGylated liposomal doxorubicin | Breast cancer | Intercalates DNA, inhibits topoisomerase II | Liposome | Intravenous | Approved (EU) | Reduced cardiotoxicity | Neutropenia, alopecia | Available in EU | - |
| **Vyxeos (CPX-351, Liposomal Daunorubicin and Cytarabine)** | EPR effect | Liposomal combination of daunorubicin and cytarabine | Acute myeloid leukemia (AML) | DNA intercalation and inhibition of DNA synthesis | Liposome | Intravenous | Approved | Improved overall survival | Febrile neutropenia, thrombocytopenia | Available | - |
| **Visudyne (Verteporfin)** | Targeted photodynamic therapy | Liposomal verteporfin | Age-related macular degeneration | Generates reactive oxygen species upon light activation | Liposome | Intravenous | Approved | Stabilization of vision loss | Visual disturbances, injection site reactions | Available | - |
| **ThermoDox (Celsion Corporation)** | Heat-triggered release | Thermosensitive liposomal doxorubicin | Liver cancer | Heat-triggered release of doxorubicin | Thermosensitive liposome | Intravenous | Phase III | Enhanced drug accumulation at tumor site | Fever, nausea | Not available | NCT00617981 |
| **Balstilimab** | Ultrasound-triggered BBB opening | Liposomal Doxorubicin | Glioblastoma | Ultrasound enhances drug penetration and activation | Liposome | Intravenous | Phase IIa | Enhanced drug delivery to brain tumors | Photosensitivity, nausea | Not available | NCT05864534 |
| **Comirnaty (Pfizer-BioNTech) and Spikevax (Moderna)** | Lipid nanoparticle encapsulation | mRNA vaccines encoding SARS-CoV-2 spike protein | COVID-19 | Induces immune response against SARS-CoV-2 | LNPs | Intramuscular | Approved | High efficacy in preventing COVID-19 | Injection site reactions, fever | Available | - |
| **Onpattro (Patisiran)** | Lipid nanoparticle delivery | siRNA targeting transthyretin (TTR) mRNA | Hereditary transthyretin-mediated amyloidosis | RNA interference reduces TTR protein production | LNPs | Intravenous | Approved | Reduction in TTR protein levels | Infusion-related reactions | Available | - |
| **Marqibo (Vincristine Liposome Injection)** | EPR effect | Liposomal vincristine | Acute lymphoblastic leukemia (ALL) | Inhibits microtubule formation | Liposome | Intravenous | Approved | Improved drug delivery to tumor cells | Peripheral neuropathy, constipation | Available | - |
| **Abraxane (Paclitaxel Nanoparticle Albumin-Bound)** | EPR effect | Albumin-bound paclitaxel nanoparticles | Breast, lung, and pancreatic cancer | Inhibits microtubule disassembly | Nanoparticle | Intravenous | Approved | Enhanced delivery and reduced toxicity | Neutropenia, neuropathy | Available | - |
| **Eteplirsen** | Targeted delivery | Phosphorodiamidate morpholino oligomer (PMO) | Duchenne muscular dystrophy | Skips exon 51 in dystrophin pre-mRNA | Lipid nanoparticle | Intravenous | Approved | Increased dystrophin production | Infusion-related reactions | Available | - |